\documentclass[twocolumn]{aastex631}

\usepackage{multirow}

\shorttitle{400-day afterglow of GRB~230815A}
\shortauthors{J.\ K.\ Leung et al., PanRadio GRB Collaboration}
\graphicspath{{./}{figures/}}

\usepackage[T1]{fontenc}
\usepackage{amsmath}

\begin{document}

\title{First Results from the PanRadio Gamma-ray Burst Collaboration:\\The 400-day Afterglow of GRB~230815A}

\correspondingauthor{James K.\ Leung}
\email{jamesk.leung@utoronto.ca}

\author[0000-0002-9415-3766]{James K.\ Leung}
\affiliation{David A. Dunlap Department of Astronomy and Astrophysics, University of Toronto, 50 St. George Street, Toronto, ON M5S 3H4, Canada}
\affiliation{Dunlap Institute for Astronomy and Astrophysics, University of Toronto, 50 St. George Street, Toronto, ON M5S 3H4, Canada}
\affiliation{Racah Institute of Physics, The Hebrew University of Jerusalem, Jerusalem 91904, Israel}

\author[0000-0001-6544-8007]{Gemma E.\ Anderson}
\affiliation{International Centre for Radio Astronomy Research, Curtin University, GPO Box U1987, Perth, WA 6845, Australia}

\author[0000-0001-9149-6707]{Alexander J.\ van der Horst}
\affiliation{Department of Physics, George Washington University, 725 21st St NW, Washington, DC 20052, USA}

\author[0009-0005-5031-8177]{Claire Morley}
\affiliation{International Centre for Radio Astronomy Research, Curtin University, GPO Box U1987, Perth, WA 6845, Australia}

\author[0000-0003-4876-7756]{Benjamin Schneider}
\affiliation{Aix Marseille Universit\'e, CNRS, CNES, LAM, 13205 Marseille Cedex 1, France}
\affiliation{Massachusetts Institute of Technology, Kavli Institute for Astrophysics and Space Research, Cambridge, MA 02139, USA}

\author[0000-0002-3137-4633]{Fabio De Colle}
\affiliation{Instituto de Ciencias Nucleares, Universidad Nacional Aut\'onoma de M\'exico, A. P. 70-543 04510 D. F. Mexico}

\author[0000-0003-4924-7322]{Om Sharan Salafia}
\affiliation{INAF - Osservatorio Astronomico di Brera, via E. Bianchi 46, I-23807 Merate (LC), Italy}
\affiliation{INFN - Sezione di Milano–Bicocca, Piazza della Scienza 3, 20126 Milano (MI), Italy}

\author[0000-0001-5876-9259]{Giancarlo Ghirlanda}
\affiliation{INAF - Osservatorio Astronomico di Brera, via E. Bianchi 46, I-23807 Merate (LC), Italy}
\affiliation{INFN - Sezione di Milano–Bicocca, Piazza della Scienza 3, 20126 Milano (MI), Italy}

\author[0000-0003-3507-335X]{Sarah I.\ Chastain}
\affiliation{Department of Physics and Astronomy, University of New Mexico, Albuquerque, NM 87131, USA}
\affiliation{Department of Physics \& Astronomy, Texas Tech University, Box 41051, Lubbock, TX, 79409-1051, USA}

\author[0000-0003-3441-8299]{Adelle J.\ Goodwin}
\affiliation{International Centre for Radio Astronomy Research, Curtin University, GPO Box U1987, Perth, WA 6845, Australia}

\author[0000-0001-8802-520X]{Ashna Gulati}
\affiliation{Sydney Institute for Astronomy, School of Physics, The University of Sydney, NSW 2006, Australia}
\affiliation{CSIRO Space and Astronomy, PO Box 76, Epping, NSW 1710, Australia}
\affiliation{ARC Centre of Excellence for Gravitational Wave Discovery (OzGrav), Hawthorn, VIC 3122, Australia}

\author[0000-0003-2705-4941]{Lauren Rhodes}
\affiliation{Trottier Space Institute at McGill, 3550 Rue University, Montreal, Quebec H3A 2A7, Canada}
\affiliation{Department of Physics, McGill University, 3600 Rue University, Montreal, Quebec H3A 2T8, Canada}

\author[0000-0003-4501-8100]{Stuart D.\ Ryder}
\affiliation{School of Mathematical and Physical Sciences, Macquarie University, NSW 2109, Australia}
\affiliation{Astrophysics and Space Technologies Research Centre, Macquarie University, Sydney, NSW 2109, Australia}

\author[0000-0001-9842-6808]{Ashley A.\ Chrimes}
\affiliation{European Space Agency (ESA), European Space Research and Technology Centre (ESTEC), Keplerlaan 1, 2201 AZ Noordwijk, The Netherlands}
\affiliation{Department of Astrophysics/IMAPP, Radboud University Nijmegen, P.O. Box 9010, Nijmegen, 6500 GL, The Netherlands}

\author[0000-0002-7320-5862]{Valerio D'Elia}
\affiliation{Space Science Data Center (SSDC) - Agenzia Spaziale Italiana (ASI), Via del Politecnico snc, 00133 Roma, Italy}

\author[0000-0002-4650-1666]{Mathieu de Bony de Lavergne}
\affiliation{Aix Marseille Universit\'e, CNRS/IN2P3, CPPM, 13288 Marseille Cedex 9, France}
\affiliation{IRFU, CEA, Universit\'e Paris-Saclay, F-91191 Gif-sur-Yvette, France}

\author[0000-0002-4036-7419]{Massimiliano De Pasquale}
\affiliation{University of Messina, MIFT Department, via F. S. D’Alcontres 31, Messina, 98166, Italy}

\author[0000-0001-7717-5085]{Antonio de Ugarte Postigo}
\affiliation{Aix Marseille Universit\'e, CNRS, CNES, LAM, 13205 Marseille Cedex 1, France}

\author[0000-0002-8028-0991]{Dieter H.\ Hartmann}
\affiliation{Department of Physics and Astronomy, Clemson University, Clemson, SC 29634-0978, USA}

\author[0000-0002-5826-0548]{Benjamin P. Gompertz}
\affiliation{School of Physics and Astronomy, University of Birmingham, Birmingham B15 2TT, UK}
\affiliation{Institute for Gravitational Wave Astronomy, University of Birmingham, Birmingham B15 2TT}

\author[0000-0001-7821-9369]{Andrew J.\ Levan}
\affiliation{Department of Astrophysics/IMAPP, Radboud University Nijmegen, P.O. Box 9010, Nijmegen, 6500 GL, The Netherlands}
\affiliation{Department of Physics, University of Warwick, Coventry, CV4 7AL, UK}

\author[0000-0002-2686-438X]{Tara Murphy}
\affiliation{Sydney Institute for Astronomy, School of Physics, The University of Sydney, NSW 2006, Australia}
\affiliation{ARC Centre of Excellence for Gravitational Wave Discovery (OzGrav), Hawthorn, VIC 3122, Australia}

\author[0000-0002-9516-1581]{Gavin P.\ Rowell}
\affiliation{School of Physics, Chemistry and Earth Sciences, The University of Adelaide, Adelaide SA 5005, Australia}

\author[0000-0002-7930-2276]{Thomas D.\ Russell}
\affiliation{INAF, Istituto di Astrofisica Spaziale e Fisica Cosmica, Via U. La Malfa 153, I-90146 Palermo, Italy}

\author[0000-0003-1500-6571]{Fabian Sch\"ussler}
\affiliation{IRFU, CEA, Universit\'e Paris-Saclay, F-91191 Gif-sur-Yvette, France}

\author[0000-0001-5803-2038]{Rhaana L.\ C.\ Starling}
\affiliation{School of Physics and Astronomy, University of Leicester, University Road, Leicester LE1 7RH, UK}

\author[0000-0003-3274-6336]{Nial R.\ Tanvir}
\affiliation{School of Physics and Astronomy, University of Leicester, University Road, Leicester LE1 7RH, UK}

\author[0000-0001-9398-4907]{Susanna D.\ Vergani}
\affiliation{LUX, Observatoire de Paris, PSL University, CNRS, Sorbonne University, 92190 Meudon, France}

\author[0000-0002-3101-1808]{Ralph A.M.J. Wijers}
\affiliation{Anton Pannekoek Institute for Astronomy, University of Amsterdam, Science Park 904, P.O. Box 94249, 1090 GE Amsterdam, The Netherlands}

\begin{abstract}
We introduce the PanRadio Gamma-ray Burst (GRB) program carried out on the Australia Telescope Compact Array: a systematic, multi-year, radio survey of all southern \textit{Swift} GRB events, comprehensively following the multi-frequency evolution of their afterglows from within an hour to years post-burst. 
We present the results of the 400-day observing campaign following the afterglow of the long-duration (collapsar) GRB~230815A, the first one detected through this program. 
Typically, GRB~230815A would not otherwise receive traditional radio follow-up, given it has no known redshift and lacks comprehensive multi-wavelength follow-up due to its high line-of-sight extinction with $A_V=2.3$. 
We found its early X-ray jet break at ${\sim}0.1$ days post-burst to be at odds with the evolution of the multi-frequency radio light curves that were traced over an unusually long duration of $400$ days. 
The radio light curves approximately evolved (with minor deviations) according to the standard self-similar expansion for a relativistic blast wave in a homogeneous environment prior to the jet break, showing no evidence for evolutions of the microphysical parameters describing the electron acceleration processes. 
We reconcile these features by proposing a two-component jet: the early X-ray break originates from a narrow component with a half-opening angle ${\sim}2.1\degr$, while the evolution of the radio afterglow stems from a wider component with a half-opening angle of $\gtrapprox 35\degr$.
The PanRadio GRB program will establish a sample of comprehensively followed GRBs, where a rigorous inspection of their microphysical and dynamical parameters can be performed, thereby revealing the diversity of features in their outflows and environments. 
\end{abstract}


\section{Introduction} \label{sec:intro}

Ultra-relativistic collimated outflows are thought to produce gamma-ray bursts (GRBs), which for a brief moment, radiate enough energy to outshine the rest of the Universe. 
These outflows are launched by a newborn, rapidly spinning, compact object following the collapse of a massive star or the merger of a compact binary \citep{Paczynski1986,Rees1992,Woosley1993,Meszaros1997}. 
As the outflow expands ballistically within the circumburst medium (CBM) that surrounds the progenitor, it gives rise to a ``forward'' shock, propagating outward into the CBM. 
When the shock sweeps up a sufficient amount of mass, the ejecta significantly decelerate and drive a ``reverse'' shock, propagating toward the inner layers of the outflow in the comoving frame of the blast wave. 
The diffusive acceleration of electrons within these shocks leads to synchrotron and inverse Compton emission, producing a GRB ``afterglow'' detectable from radio to very-high-energy (TeV) gamma rays \citep[e.g.,][]{Costa1997,Frail1997,vanParadijs1997,HESS2019,MAGIC2019}. 

As sites of particle acceleration, the forward and reverse shocks produce two distinct synchrotron components detectable at radio (gigahertz) frequencies, hence a forward-reverse shock model is needed for interpreting the early afterglow. 
The broadband spectrum of each synchrotron component is described by the peak flux density and the characteristic break frequencies, which all evolve as a function of time \citep{Sari1998,Panaitescu2000,Granot2002,Gao2013}. 
The observed spectral evolution can then be related to the physical properties describing the system, such as the energetics of the outflow, the density and structure of the CBM (and, by extension, the mass-loss history of the progenitor in the case of GRBs originating from collapsars), and the shock microphysics (i.e., the fraction of energy in the magnetic fields and emitting electron population, as well as the Lorentz factor distribution). 
While the forward shock emission can take tens to hundreds of days to peak at gigahertz frequencies \citep{Chandra2012}, the reverse shock often peaks and fades away within a day to a few days post-burst \citep[e.g.,][]{Anderson2014,Anderson2018}. 
This highlights the need for high-cadence multi-frequency radio observations from early (less than a day post-burst) through to late times (more than a year post-burst) to properly constrain and disentangle the properties of the reverse and forward shock components. 

Comprehensive radio monitoring of an afterglow will therefore enable a complete and rigorous inspection of the dynamical and shock parameters that describe the blast wave and its environment. 
Despite GRBs being the most powerful explosions in the Universe, we still lack a complete picture of the generation and properties of the outflows and how they impact their surrounding environment. 
This is in large part due to the limited radio observations of GRBs. 
It is both difficult to rapidly schedule radio observations following the appearance of a new GRB event and difficult to follow the radio afterglows up until late times (more than a year post-burst) due to the limited number of radio interferometer arrays in the world with sufficient sensitivity for this task; for a brief review of radio afterglow observations in the first 14 years since the discovery of the afterglow (1997 to 2011), see \citet[][]{Chandra2012}. 

The GRBs that have radio follow-up in the literature have often been selectively targeted based on known properties --- such as the existence of an optical and/or X-ray counterpart, a low redshift, a high gamma-ray fluence \citep[e.g., in the case of GRB~221009A;][]{Bright2023,Laskar2023,Rhodes2024}, those suspected to be optically ``dark'' bursts \citep[e.g.,][]{Djorgovski2001,Jakobsson2004,vanderHorst2009}, candidate lensing events \citep[e.g.,][]{Giarratana2023,Leung2026}, and those suspected to be at high redshift \citep[e.g.,][and references therein]{Salvaterra2015,Fausey2025} --- which leads to inherent biases in our understanding of their phenomenological and physical properties.  
Of those that have radio follow-up, an even smaller subset are followed to very late times. 
Indeed, in the three decades since the discovery of the first afterglow, only approximately 10 GRBs had comprehensive multi-wavelength coverage extending to more than a year post-burst 
(e.g., GRBs 970508, 980703, 000418, 030329, 130427A, 170817A, 171205A, 190829A, 221009A; \citealt{Frail2000,Frail2003,Berger2001,vanderHorst2008,vanderHorst2014,Makhathini2021,Leung2021,Maity2021,Rhodes2020,Salafia2022a,Laskar2023,Rhodes2024}).

As a result, the transition of the blast wave to the non-relativistic regime and the deep-Newtonian regime, often occurring years post-burst, is not very well constrained for most GRBs. 
In the non-relativistic regime, the blast wave has decelerated to sub-relativistic Lorentz factors and becomes quasi-spherical, where the dynamics can be approximately described by the Sedov-von~Neumann-Taylor self-similar solution \citep[e.g.,][]{Frail2000,Livio2000,Zhang2009,DeColle2012,vanEerten2012}. 
At even later times, when the bulk of the shocked electrons become non-relativistic, the blast wave enters the deep-Newtonian regime, resulting in  a shallower decay than that predicted in the non-relativistic regime \citep{Sironi2013}.
Radio monitoring of the GRB after the blast wave has entered the non-relativistic and deep-Newtonian phase is important because it allows for accurate calorimetry regardless of the initial jet structure \citep{Frail2000}. 
Calorimetric calculations have only been performed in a few works and for a small sample of bursts \citep[e.g.,][]{Frail2000,Frail2005,vanderHorst2008,Shivvers2011} so a precise distribution of GRB energy budgets is still lacking. 
Additionally, the tracing of the radio spectral evolution into late times allows for the density profile of the CBM to be mapped to a larger radius (and farther back in time pre-explosion), revealing information about both the host galaxy environment and mass-loss history of the progenitor. 
For instance, changes in the density profile are expected at the wind-termination-shock radius \citep[e.g.,][]{Peer2006, Gendre2007,Schulze2011} -- the interface between the progenitor stellar winds and the surrounding interstellar medium (ISM). 
The transition could explain why some long GRBs are better explained by an ISM environment \citep[a dense ISM may suppress the propagation of the stellar wind; see, e.g.,][and references therein]{Chrimes2022}; however, no observations to date have convincingly shown this transition. 
A recent probe of the CBM environment at very late times for GRB~171205A found that even at over $1\,000$ days post-burst, the afterglow was consistent with the burst exploding into a stellar wind environment with no signs of any regime transition \citep{Leung2021,Maity2021}. 
Following events to these very late times at higher energies (than radio) is difficult but not unprecedented; e.g., the X-ray afterglow of GRB~130427A was followed for ${\sim} 80$\,Ms (or ${\sim} 1\,000$\,days) and similarly showed an unperturbed decay across 3 orders of magnitude in time \citep{DePasquale2016}. 

These results (or lack thereof) motivated the Panoptic Radio View of Gamma-ray Bursts (``PanRadio GRB'') program -- a comprehensive radio follow-up program of GRBs. 
The PanRadio GRB program will provide a more complete, more unbiased, multi-frequency ($1-50$\,GHz), and high-cadence radio view of GRBs from early ($<1$ day post-burst) through to late ($>1$ year post-burst) times. 
The aim of the program is to ultimately determine the true prevalence of radio afterglow emission by more than doubling the number of comprehensively modelled multi-wavelength afterglows, allowing for a sample analysis probing how the physical properties of GRB outflows evolve with time.  
In this paper, we present our first results: a 400-day observing campaign following the afterglow of GRB~230815A, which was the first afterglow detected through the program. 
In Section~\ref{sec:230815a-data}, we introduce the PanRadio GRB program and present the data obtained from our multi-wavelength observing campaign for GRB~230815A. 
In Section~\ref{sec:results}, we provide a description of the basic features we observe in the GRB~230815A afterglow.
In Section~\ref{sec:discussion}, we provide a unified physical interpretation of the afterglow, make inferences about the dynamical and microphysical evolution of the afterglow, and discuss the outlook for the broader PanRadio GRB program. 
We finally present our conclusions in Section~\ref{sec:conclusions}. 

Throughout this paper, we assume a flat $\Lambda$CDM cosmology with $H_0=~69.6$\,km\,s$^{-1}$\,Mpc$^{-1}$, $\Omega_{\text{M}} = 0.286$, and $\Omega_{\Lambda} = 0.714$ (these were chosen to match the cosmological parameters used in \citet{Duncan2023}, which we compare some of our results in this manuscript with). 
We assume the following convention throughout, unless specified otherwise, for the temporal index $\alpha$ and spectral index $\beta$: $S(t, \nu) \propto t^\alpha \nu^\beta$, where $S(t,\nu)$ is the flux density as a function of time and frequency, respectively. 
We further define the X-ray photon index $\Gamma$ to be related to the spectral index by $\beta_{\rm x} \equiv 1 - \Gamma$.
All errors are given at the 68\% confidence level. 

\section{Observations and Data Reduction} \label{sec:230815a-data}

\subsection{PanRadio GRB Program}

The PanRadio GRB program on the Australia Telescope Compact Array (ATCA) began in 2023 April.  
Through the program, we have and will continue to obtain high-cadence follow-up across a broad frequency range ($1-50$\,GHz) for a large sample of \textit{Swift}-localised GRBs observed over a period of at least 3 years, allowing us to build a more unbiased early-to-late-time radio sample of GRBs. 
Ultimately, we will use this sample to improve our understanding of open questions relating to the dynamics and shock physics of GRBs -- in particular, regarding the composition and morphology of GRB jets; whether there is universality in some microphysical shock parameters among all GRBs as well as constraints on the evolution of these microphysical parameters as a function of time; and the implications of the inferred total energy budgets and environmental properties on the diversity of GRB progenitors. 

A follow-up paper (Anderson et al., \textit{in preparation}) will provide a complete description of the survey, the observing strategy, and initial sample analyses from bursts analysed in the first 2 years of the program. 
The remainder of this paper describes the 400-day observing campaign following the evolution of the afterglow for GRB~230815A; as the first afterglow detected through the PanRadio GRB program, we use it to highlight the potential science outcomes we can achieve with the entire sample following the completion of our program. 
Specifically, we will show through this campaign how the comprehensive early-to-late-time follow-up of GRBs under the PanRadio GRB program can directly lead to insights into the electron acceleration processes, density profiles, and jet properties of the wider GRB population. 

\subsection{GRB~230815A Observing Campaign}

GRB~230815A was detected by the \textit{Swift}/Burst Alert Telescope (BAT) at $t_0$=2023-08-15T10:49:55 UT \citep{KlinglerGCN34434}. 
Although initially reported with a $T_{90}$ duration of ${\sim} 2$\,s, the refined analysis in the $15 - 350$\,keV energy band was refined to $17.00 \pm 7.62$\,s \citep{Laha2023GCN34449}. 
The GRB was also detected by the \textit{Fermi}/Gamma-ray Burst Monitor (GBM) \citep{MailyanGCN34440} and its $T_{90}$ duration in the $50-300$\,keV band was 5.2\,s, with the peak energy $E_{\rm peak} = 235 \pm 25$\,keV in the time-averaged spectrum from $t_0+0.002$ to $t_0+3.264$\,s from a fit with the Band function. 
The prompt emission properties suggest GRB~230815A is likely a standard long-soft GRB with a collapsar origin. 
The \textit{Swift}/XRT identified the X-ray afterglow, localising it to RA (J2000)=12:18:53.34, Dec (J2000): -58:03:10.4, with a positional uncertainty of 2\farcs2 \citep{BeardmoreGCN34437}. 
The optical wavelengths were suppressed by considerable line-of-sight Galactic extinction, with a reddening of $E(B-V) = 0.73$, corresponding to a total visual extinction of $A_V = 2.3$ for the typical Milky Way extinction-to-reddening ratio $R_V=3.1$ \citep{Schlegel1998}.
This ultimately led to non-detections in all \textit{Swift}/Ultra-violet Optical Telescope (UVOT) \citep{BreeveldGCN34446} and Southern Astrophysical Research (SOAR)/Goodman \citep{KilpatrickGCN34452} filters, limiting follow-up with optical facilities and preventing a redshift determination. 

At longer wavelengths, our team was able to detect and follow the evolution of the radio \citep{Leung2023GCN34518} and near-infrared \citep{SchneiderGCN34467} afterglow with ATCA and the Very Large Telescope High Acuity Wide-field $K$-band Imager (VLT/HAWK-I), respectively. 
We describe the observing campaign and the data reduction process for our radio and near-infrared observations below. 

Our full set of measurements can be found in Tables~\ref{tab:radio} and \ref{tab:optical}. 
Table~\ref{tab:radio} shows the measurements from our radio observing campaign. 
Table~\ref{tab:optical} shows the optical and near-infrared measurements, including the upper limits acquired from the GCN Circulars as well as our near-infrared observing campaign.
In this work, we used the \textit{Swift}/XRT products, which were produced via the automated \textit{Swift}/XRT analysis tools \citep{Evans2009} and are publicly available through the UK \textit{Swift} Science Data Centre.\footnote{\url{https://www.swift.ac.uk/xrt_spectra/01185505/}} 
The table of unabsorbed \textit{Swift}/XRT fluxes at $0.3 - 10$\,keV was taken from the Swift Burst Analyser\footnote{\url{https://www.swift.ac.uk/burst_analyser/01185505/}} \citep{Evans2010}; we considered data from the windowed timing (WT) and photon counting (PC) modes but excluded all slewing data.
The X-ray data were corrected for absorption, using 
$N_\textrm{H} = 5.64 \times 10^{21}\,\textrm{cm}^{-2}$ at $z=0$, where $N_\textrm{H} \ \textrm{(Galactic)} = 4.94 \times 10^{21}\,\textrm{cm}^{-2}$ and $N_\textrm{H} \ \textrm{(intrinsic)} = 0.7 \times 10^{21}\,\textrm{cm}^{-2}$ as obtained from the late-time spectrum using photons after 6\,ks post-burst. 

\begin{table*}
\fontsize{10pt}{16pt}\selectfont 
\centering
\caption{
ATCA radio flux-density measurements for GRB~230815A. 
Columns 1 through 9 show the number of days since the \textit{Swift}/BAT trigger (10:49:55 UT on 2023 August 15), the array configuration, and the flux-density measurements (or $3\sigma$ limit for non-detections, where $\sigma$ here is simply the rms noise) in \textmu Jy at 2.1, 5.5, 9, 16.7, 21.2, 34, and 44\,GHz. 
}
\label{tab:radio}
\begin{tabular}{ccccccccc}
\hline\hline
dT & Config 
& $S_{2.1\,\textrm{GHz}}$ 
& $S_{5.5\,\textrm{GHz}}$
& $S_{9\,\textrm{GHz}}$
& $S_{16.7\,\textrm{GHz}}$
& $S_{21.2\,\textrm{GHz}}$
& $S_{34\,\textrm{GHz}}$
& $S_{44\,\textrm{GHz}}$
\\
(days) & 
& (\textmu Jy)
& (\textmu Jy)
& (\textmu Jy)
& (\textmu Jy)
& (\textmu Jy)
& (\textmu Jy)
& (\textmu Jy)
\\
\hline
\phantom{0}1.81 & 6D & --- & $45 \pm 19$ & $83 \pm 19$  & $<288$ & $<597$ & --- & --- \\
\phantom{0}3.63 & 6D & --- & $49 \pm 17$ & $76 \pm 16$ & --- & --- & --- & $168 \pm 38\phantom{0}$ \\
\phantom{0}7.87 & 6D & --- & $48 \pm 14$ & $102 \pm 17$ & --- & --- & --- & --- \\
\phantom{0}9.60 & 6D & --- & $<39$ & $79 \pm 15$ & $210 \pm 25\phantom{0}$ & $223 \pm 41\phantom{0}$ & --- & --- \\
25.49 & H168 & ---  & $<75$ & $118 \pm 21\phantom{0}$ & $369 \pm 34\phantom{0}$ & $475 \pm 52\phantom{0}$ & --- & --- \\
39.46 & H168 & --- & $<93$ & $175 \pm 23\phantom{0}$ & $502 \pm 45\phantom{0}$ & $598 \pm 58\phantom{0}$ & $410 \pm 57\phantom{0}$ & --- \\
61.45 & 750B & --- & $<72$ & $134 \pm 27\phantom{0}$ & $297 \pm 33\phantom{0}$ & $424 \pm 58\phantom{0}$ & $307 \pm 47\phantom{0}$ & --- \\
96.45 & H214 & --- & $<84$ & $<99$ & $313 \pm 86\phantom{0}$  & $<498$ & $<636$ & --- \\
129.45\phantom{0} & 6D & $<129$ & $107 \pm 22\phantom{0}$ & $117 \pm 19\phantom{0}$ & $115 \pm 33\phantom{0}$ & $160 \pm 48\phantom{0}$ & --- & --- \\
212.99\phantom{0} & 6A & $<108$ & $<57$ & $112 \pm 18\phantom{0}$ & $84 \pm 24$ & $<174$ & --- & --- \\
302.83\phantom{0} & 6D & --- & $84 \pm 19$ & $47 \pm 14$ & $57 \pm 22$ & $<75$ & --- & --- \\
405.77\phantom{0} & 6A & --- & $80 \pm 17$ & $49 \pm 9\phantom{0}$ & --- & --- & --- & --- \\
\hline
\end{tabular}

\end{table*}

\begin{table*}[h!] 
\centering
\caption{
Optical and near-infrared measurements for GRB~230815A. 
Columns 1 through 8 show the number of days since the \textit{Swift}/BAT trigger (10:49:55 UT on 2023 August 15), the observing instrument, the photometric filter, the effective frequency corresponding to the photometric filter, the magnitude measurement, the magnitude measurement converted to a flux density, the magnitude measurement converted to a flux density \textit{with} correction for Galactic extinction, and the reference from where the measurement was obtained. 
Limits are derived by performing forced aperture photometry at the site of the source. 
The correction for Galactic extinction in the various filters is based on the dust maps from \citet{Schlegel1998}, which were accessed through \url{https://irsa.ipac.caltech.edu/applications/DUST/}; for the \textit{Swift}/UVOT filters, the extinction curves were determined using the values in \citet{Roming2009}.
}
\label{tab:optical}
\begin{tabular}{cccccccccc}
\hline\hline
dT (days) & Instrument & Filter & $\nu$ (Hz) & Magnitude & $S_\nu$ (\textmu Jy) & $S_\nu$ (\textmu Jy) & References \\
&&&& \textit{uncorrected} & \textit{uncorrected} & \textit{corrected} &  \\
\hline
0.011 & \textit{Swift}/UVOT & $v$ & 5.54e+14 & $>18.50$ & $<144.54$ & $<894.14$ & \citealt{BreeveldGCN34446}\\
0.011 & \textit{Swift}/UVOT & $uvw2$ & 1.44e+15 & $>18.80$ & $<109.65$ & $<13711.08$ & \textquotedbl\textquotedbl\\
0.012 & \textit{Swift}/UVOT & $b$ & 6.90e+14 & $>19.30$ & $<69.18$ & $<766.97$ &  \textquotedbl\textquotedbl\\
0.012 & \textit{Swift}/UVOT & $uvw1$ & 1.12e+15 & $>18.60$ & $<131.83$ & $<6529.31$ & \textquotedbl\textquotedbl\\
0.014 & \textit{Swift}/UVOT & $uvm2$ & 1.34e+15 & $>18.30$ & $<173.78$ & $<42761.16$ & \textquotedbl\textquotedbl\\
0.039 & \textit{Swift}/UVOT & $u$ & 8.51e+14 & $>19.40$ & $<63.10$ & $<1147.47$ & \textquotedbl\textquotedbl\\
0.519 & SOAR/Goodman & $i$ & 3.92e+14 & $>20.50$ & $<22.91$ & $<72.44$ & \citealt{KilpatrickGCN34452} \\ 
0.52 & VLT/HAWK-I & $H$ & 1.87e+14 & $20.19 \pm 0.02$ & $30.48 \pm 0.56$ & $41.65 \pm 0.77$ & this work \\
0.542 & SOAR/Goodman & $z$ & 3.09e+14 & $>17.90$ & $<251.19$ & $<591.56$ & \citealt{KilpatrickGCN34452}\\
1.52 & VLT/HAWK-I & $H$ & 1.87e+14 & $22.10 \pm 0.08$ & $\phantom{0}5.25 \pm 0.39$ & $\phantom{0}7.84 \pm 0.58$ & this work \\
219.80 & VLT/HAWK-I & $H$ & 1.87e+14 & $>24.50$ & $<0.58$ & $<0.79$ & \textquotedbl\textquotedbl\\
\hline
\end{tabular}
\end{table*}

\subsubsection{ATCA}
We observed GRB~230815A under the PanRadio GRB program (PI: Anderson; C3542). 
Our first ATCA observations of GRB~230815A started at 02:00 UT on 2023 August 17 (or 1.6 days post-burst), leading to the discovery of the radio counterpart \citep{Leung2023GCN34518}.  
We note that this GRB did not trigger the rapid-response capabilities of the telescope (which would carry out observations $< 1$ day post-burst) because a message containing the \textit{Swift}/BAT position was not circulated in the standard alerts stream, which was unusual\footnote{This has only occurred 15 times since the launch of \textit{Swift}; see: \url{https://gcn.gsfc.nasa.gov/swift_grbs.html}} for events with a \textit{Swift}/BAT detection. 
Since then, we have collected an additional 11 epochs of data, with the final epoch occurring  on 2024 September 24 (406 days post-burst). 

Across the ATCA observing campaign, we used the 2.1, 5.5/9, 16.7/21.2, 33/35, and 43/45\,GHz receiver configurations. 
We reduced the visibility data from each observation using standard {\sc Miriad} procedures \citep{Sault1995}. 
We used B1934$-$638 to set the flux-density scale and B1129$-$58 to calibrate the time-variable complex gains. 
The singular exception is for the 9\,GHz data obtained on 2024 June 13 UT, where the bootstrapping of the flux-density scale to B1934$-$638 failed; in this case, the flux-density scale was instead bootstrapped to B1129$-$58,\footnote{We are able to bootstrap the flux-density scale to B1129$-$58 because this source exhibits very limited variability at 9\,GHz historically: \url{https://www.narrabri.atnf.csiro.au/calibrators/calibrator_database_viewcal?source=1129-58}} contributing to an additional uncertainty (of 15\%, from the historical variability observed in the source) added in quadrature to the statistical uncertainties in the flux-density measurement. 
We also used B1934$-$638 to determine the bandpass response at the 2.1, 5.5, and 9\,GHz frequency bands; at higher-frequency bands (16.7\,GHz and higher), we instead used B0727$-$115, B0537$-$441, or B1921$-$293, depending on which was higher in elevation at the start of our observing run. 
For observations at the higher-frequency bands, we accounted for, to first order, the spectral shape of the bandpass calibrators (whose spectral shape is unknown, unlike B1934$-$638) by fitting a flux-density model across the intermediate frequencies (e.g., 16.7 and 21.2 GHz). 

After calibration, we imaged the visibilities data for the target field using the {\sc Miriad} tasks {\sc invert}, {\sc clean}, and {\sc restor}, with the multi-frequency synthesis CLEAN algorithm \citep{Hogbom1974, Clark1980, Sault1994} used for deconvolution. 
The 2.1, 5.5, 9, 16.7, and 21.2\,GHz data were imaged using a 2\,GHz bandwidth, while the higher frequency data (33/35 and 43/45\,GHz) were imaged with a larger 4\,GHz bandwidth by combining near-contiguous intermediate frequency bands to improve the image sensitivity. 
For each observation, we measure the flux density of a detected source by fitting a point-source model to the restored image using the {\sc Miriad} task {\sc imfit} and report a non-detection using the rms sensitivity obtained from the residual image. 
The errors reported are purely statistical, as the systematic errors are expected to be much smaller \citep[$\lesssim 5$ per cent; e.g.,][]{Reynolds1994,Tingay2003}. 
The final set of observation details and measurements is given in Table~\ref{tab:radio}. 

\subsubsection{VLT/HAWK-I}

We carried out observations of the GRB~230815A field with the 8.2\,m VLT on Cerro Paranal (Chile).
The observations were made with the HAWK-I near-infrared camera mounted on the Unit Telescope 4 (UT4, Yepun) in the $H$-band at three epochs (0.52, 1.52, and 219.80 days post-burst) under the European Southern Observatory (ESO) program IDs 110.24CF.014 and 110.24CF.019 (PIs: N.~Tanvir, D.~Malesani, and S.~Vergani). 
The observations consisted of 20\,min acquisitions (i.e., the total on-source exposure time without overhead) for the first two epochs and 36\,min for the last epoch. 
All were obtained under excellent conditions (seeing $<0\farcs8$). 
The images were reduced using the standard ESO Reflex pipeline \citep{Freudling2013} and the certified calibration files of the night provided by ESO. 
The photometric calibration was performed using nearby stars from the Two Micron All Sky Survey catalogue \citep{Skrutskie2006}. 
We derived the magnitudes by aperture photometry using the \textsc{Astropy Photutils} package \citep[v2.0.2,][]{Bradley2024} and corrected for Galactic extinction using dust maps from \citet{Schlegel1998}, which were accessed through \url{https://irsa.ipac.caltech.edu/applications/DUST/}.

In the first two epochs (0.52 and 1.52 days post-burst), two sources consistent with the XRT error circle were detected. 
One of them showed a clear sign of fading between the two epochs, from $19.87 \pm 0.02$ to $21.77 \pm 0.08$\,AB magnitude (i.e., fading of ${\sim} 2$\,mag), confirming the variable nature of the source and was proposed as the near-infrared afterglow counterpart of the GRB \citep{SchneiderGCN34467}. 
A late-time observation was obtained at 219.80 days post-burst to investigate possible host contamination in the early epochs. 
No clear source was detected at the afterglow position down to an AB magnitude of 24.2, confirming negligible host contamination in the first epochs.
We provide the observational details and measurements in Table~\ref{tab:optical}. 

Using our VLT/HAWK-I detection at 0.52 days post-burst (corrected for Galactic extinction), along with \textit{Swift} XRT photons from 0.464 days post-burst (at 1\,keV), we assessed whether GRB~230815A could be an optically ``dark'' GRB -- i.e., whether its optical afterglow is suppressed with respect to other wavelengths due to dust obscuration from its host galaxy --
using the criterion outlined in \citet{vanderHorst2009}: $\beta_{\rm ox} < \beta_{\rm x} - 0.5$, where $\beta_{\rm ox}$ is the optical-to-X-ray spectral index, $\beta_{\rm x}$ is the X-ray spectral index, and here (only in this section for consistency with the conventions of \citet{vanderHorst2009}) the spectral index is defined as $S_\nu \propto \nu^{-\beta}$. 
We calculated $\beta_{\rm ox} = 0.74 \pm 0.03$ and, using the photon index from the PC segment of the spectrum $\Gamma = 1.94^{+0.11}_{-0.10}$ (see Section~\ref{ssec:bbsed}), obtained $\beta_{\rm x} = 0.94^{+0.11}_{-0.10}$. 
We therefore deem it very unlikely that GRB~230815A was a dark GRB and suggest instead that the majority of its optical suppression is due to Galactic extinction. 

\newpage 
\section{Results} \label{sec:results}

Here, we report and characterise the basic features of the afterglow identified from the X-ray, near-infrared, and radio observations. 

\subsection{Break in the X-ray light curve}\label{ssec:xraybreak}

\begin{figure}
    \centering
    \includegraphics[width=0.94\linewidth]{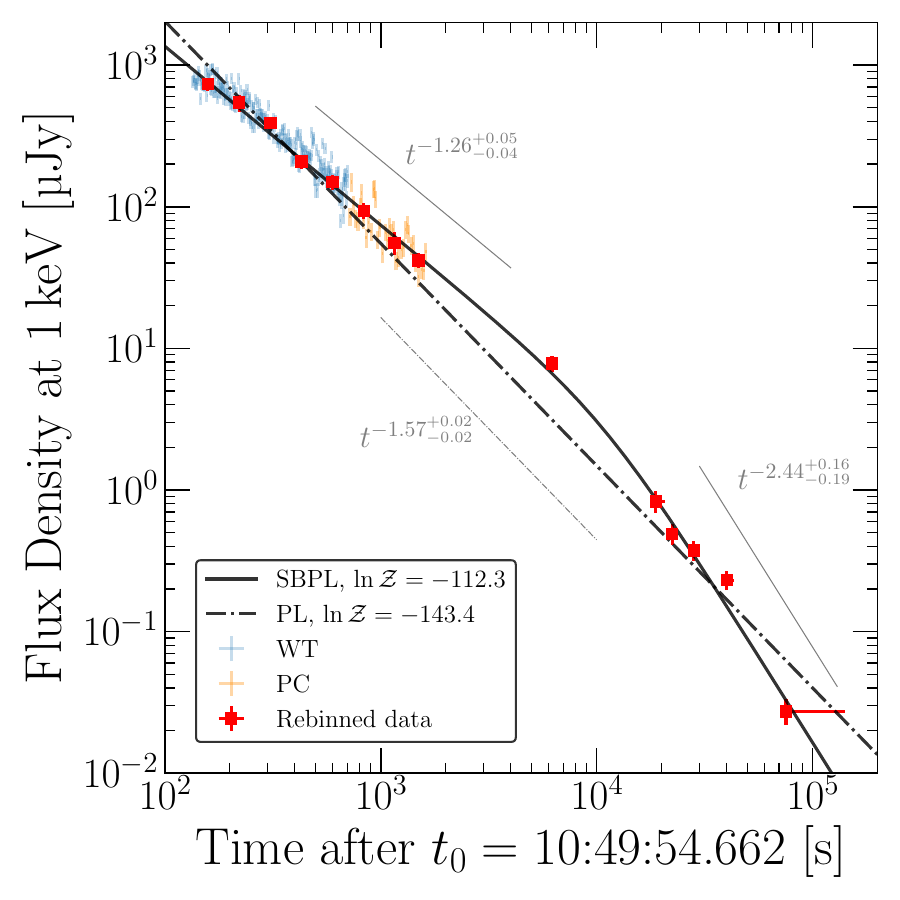}
    \caption{X-ray light curve for GRB~230815A. 
    The data points obtained by \textit{Swift}/XRT in WT mode are represented by the blue error bars while those obtained in PC mode are represented by the orange error bars.
    The data points used in the fits to the X-ray light curve, obtained from binning the \textit{Swift}/XRT data points, are represented by the red square markers. 
    Fits to a smoothly broken power law and a power law are shown by the solid and dotted-dashed lines, respectively. 
    The slopes of the fits are annotated beside the lines. 
    }
    \label{fig:xrt}
\end{figure}

\begin{table*}
\fontsize{10pt}{12pt}\selectfont
\centering
\caption{The nested sampling best-fit parameters for the smoothly broken power-law (SBPL) and power-law (PL) models describing the \textit{Swift}/XRT light curve at 1\,keV for GRB~230815A. 
Column 1 shows the model used to fit the data, columns 2 through 6 show the best-fit parameters and their uncertainties for the model fits (i.e., the median and the 16$^{\text{th}}$/84$^{\text{th}}$ percentile of the posterior distribution), column 7 shows the log evidence of the models, and column 8 shows the log Bayes factor comparing the SBPL and PL models. 
}
\label{tab:xrtlc_fits}
\begin{tabular}{|l|ccccc|c|c|}
\hline\hline
& $S_{\rm break}$ or $S_0$ (\textmu Jy) & $t_{\rm b}$ (s) & $\delta_1$ & $\delta_2$ & $s$ & $\ln \mathcal{Z}$ & $\ln \mathcal{B}_{\rm SBPL, PL}$ \\
\hline
SBPL & $2.60^{+1.53}_{-1.02}$ & $10930^{+3175}_{-2575}$ & $-1.26^{+0.05}_{-0.04}$ & $-2.44^{+0.16}_{-0.19}$ & $2.06^{+0.65}_{-0.77}$ & $-112.35$ & \multirow{2}{*}{$31.04$} \\
PL & $1.49^{+0.09}_{-0.08}$ & -- & $-1.57^{+0.02}_{-0.02}$ & -- & -- & $-143.39$ & \\
\hline
\end{tabular}
\end{table*}

The first feature we identified from the X-ray light curve is an apparent break at approximately 10\,ks post-burst as shown in Figure~\ref{fig:xrt}. 
We consider whether the presence of the break is significant (and quantify the pre/post-break decay slopes) by comparing a fit of the data to (i) a smoothly broken power-law model:
\begin{equation}
    \label{eq:sbpl}
    S_t = S_{\rm{break}}
    \Bigg[\frac{1}{2}\bigg(\frac{t}{t_{\rm b}}\bigg)^{-s\delta_1} + \frac{1}{2} \bigg(\frac{t}{t_{\rm b}}\bigg)^{-s\delta_2}\Bigg]^{-1/s},
\end{equation}
where $t$ is the time post-burst, 
$S_t$ is the flux density as a function of time $t$, 
$\delta_1$ is the pre-break decay slope, $\delta_2$ is the post-break decay slope, $S_{\rm{break}}$ and $t_{\rm{b}}$ are the approximate flux density and time post-burst corresponding to the break in the light curve, respectively, and $s$ is the smoothness parameter;
and (ii) a power-law model:
\begin{equation}
    \label{eq:pl}
    S_t = S_0 \, \left(\frac{t}{t_0}\right)^{\delta_1},
\end{equation}
where $S_0$ is the normalisation parameter at time $t_0=10$\,ks assuming a decay with a power-law index of $\delta_1$ without a break;
under a Bayesian framework. 
Due to the higher density of data points at early times compared to later times, the data points were rebinned into 20 logarithmically spaced bins, shown with the red square markers in Figure~\ref{fig:xrt}, before fitting to minimise biases in sampling the posterior distribution. 

\begin{figure*}
    \centering
    \includegraphics[width=0.94\textwidth,clip,trim={25mm 5mm 15mm 25mm}]{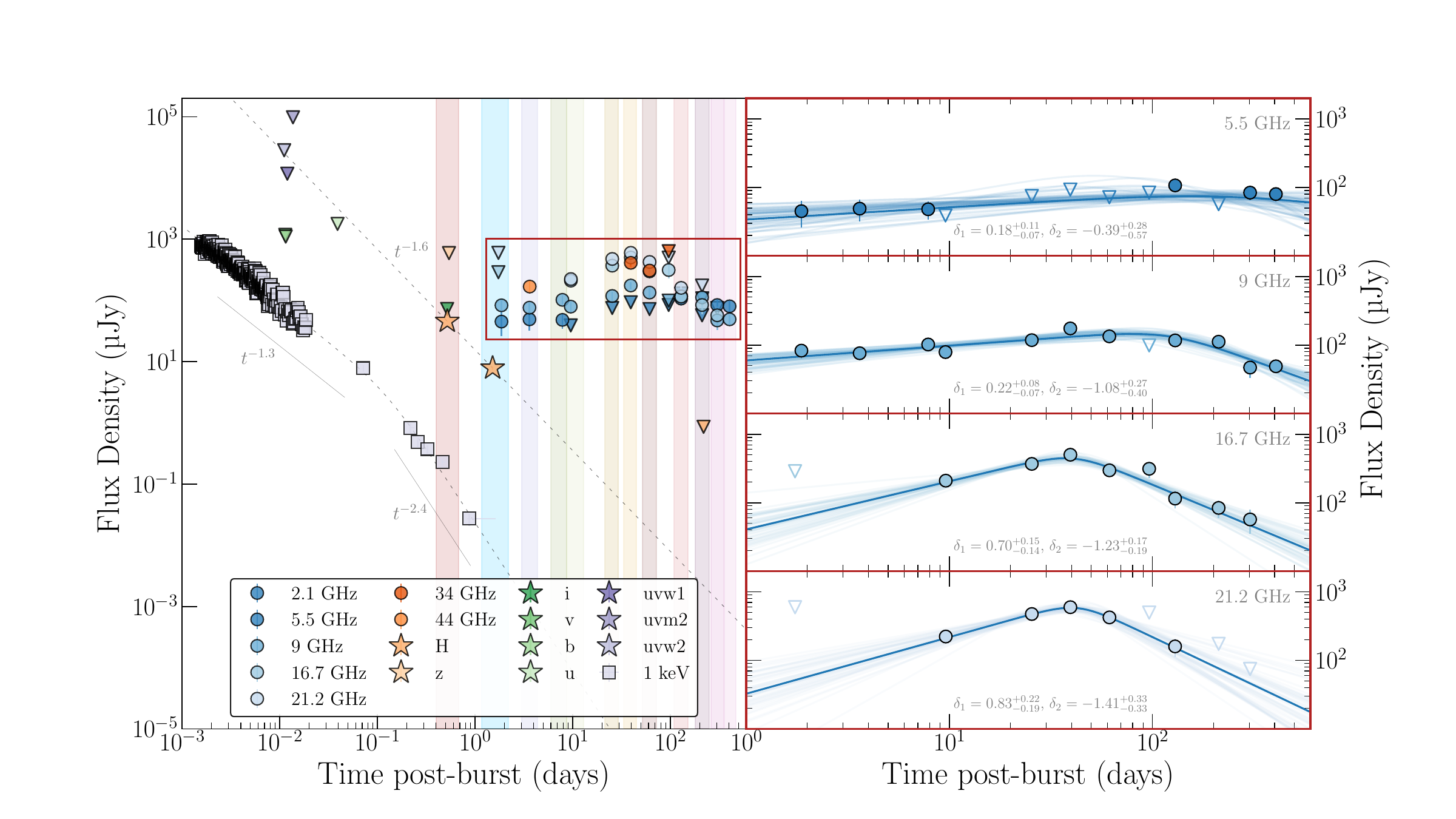}
    \caption{\textbf{\textit{Left}:} multi-wavelength light curves for GRB~230815A. 
    The radio data points are represented with circular markers, the near-infrared data points with star markers, and the X-ray data points with square markers. 
    All upper limits are shown with inverted triangular markers. 
    The fits describing the evolution of the X-ray and near-infrared light curves are shown using dashed lines, with the corresponding temporal slopes annotated next to them. 
    The radio data points surrounded by the red rectangle are shown in more detail in the \textit{right} panel. 
    The coloured vertical strips indicate the temporal windows used for constructing the SEDs presented in Figures~\ref{fig:bbsed} and \ref{fig:radioseds}, where the colour of each strip here corresponds to an SED of the same colour in those figures. 
    \textbf{\textit{Right}:} radio light curves for GRB~230815A at 5.5, 9, 16.7, and 21.2 GHz (from top to bottom). 
    Each detection is represented by a circular marker and upper limits with inverted triangular markers.
    At each frequency, the solid blue line represents the smoothly broken power-law fit (Model 1; independent $\delta_1$ and $\delta_2$ per frequency) to the light curve using the best-fit parameters estimated from nested sampling, while the 50 lines with weaker line intensity are random posterior samples used to illustrate the fit uncertainties. 
    }
    \label{fig:mwlc}
\end{figure*}

We fit the rebinned data to both models using the nested sampler {\sc Dynesty} \citep{Speagle2020} as implemented in the Bayesian inferences software {\sc Bilby} \citep{Ashton2019}. 
We performed the nested sampling using uniform priors (sampled uniformly over linear space), with 2\,000 live points, and a stopping criterion on the change in the estimated Bayesian evidence $\hat{\mathcal{Z}}$ (i.e., the normalisation integral of the posterior probability function) from one iteration to the next of $\Delta \mathrm{ln}(\hat{\mathcal{Z}}) = 0.05$. 
We show the nested sampling fits for the power-law and smoothly broken power-law models plotted on the X-ray light curve in Figure~\ref{fig:xrt} as solid and dotted-dashed lines, respectively; likewise,  the fit for the smoothly broken power-law model is also plotted in the multi-wavelength light curve in Figure~\ref{fig:mwlc}. 

The resulting posterior distributions of the parameters in the fits of the X-ray light curve to both models are reported in Table~\ref{tab:xrtlc_fits} and are also shown in Figures~\ref{fig:corner_xrt_sbpl} and \ref{fig:corner_xrt_pl} as corner plots. 
The logarithm of the Bayes factor\footnote{
The Bayes factor used for the model comparison is the ratio of the Bayesian evidence between the two models and appropriately accounts for the model complexity. The Bayesian evidence can also be viewed as the likelihood marginalised over the prior distribution. For a more complex model with more parameters, the prior volume that the likelihood is marginalised over will be larger. If the data do not require the added complexity in the model (i.e., they are ``overfit''), this will effectively downweight the high-likelihood regions of the prior volume (compared to a less complex model with a smaller prior volume), resulting in a lower Bayesian evidence, which then serves as a penalty for model complexity.} comparing the smoothly broken power-law to the power-law model $\ln \mathcal{B}_{\rm SBPL, PL} = \ln \mathcal{Z}_{\rm SBPL} - \ln \mathcal{Z}_{\rm PL} = 31.04$, where $\mathcal{Z}_{\rm model}$ is the Bayesian evidence for the data given the model, suggests the smoothly broken power law is the preferred model for explaining the X-ray data. 
The supported model indicates the presence of a break at $t_b = 10.9^{+3.2}_{-2.6}$\,ks post-burst with pre-break and post-break decays going as $t^{-1.26^{+0.05}_{-0.04}}$ and $t^{-2.44^{+0.16}_{-0.19}}$, respectively. 

\subsection{Spectral break between the X-ray and near-infrared}\label{ssec:bbsed}

\begin{figure}
    \centering
    \includegraphics[width=0.94 \linewidth]{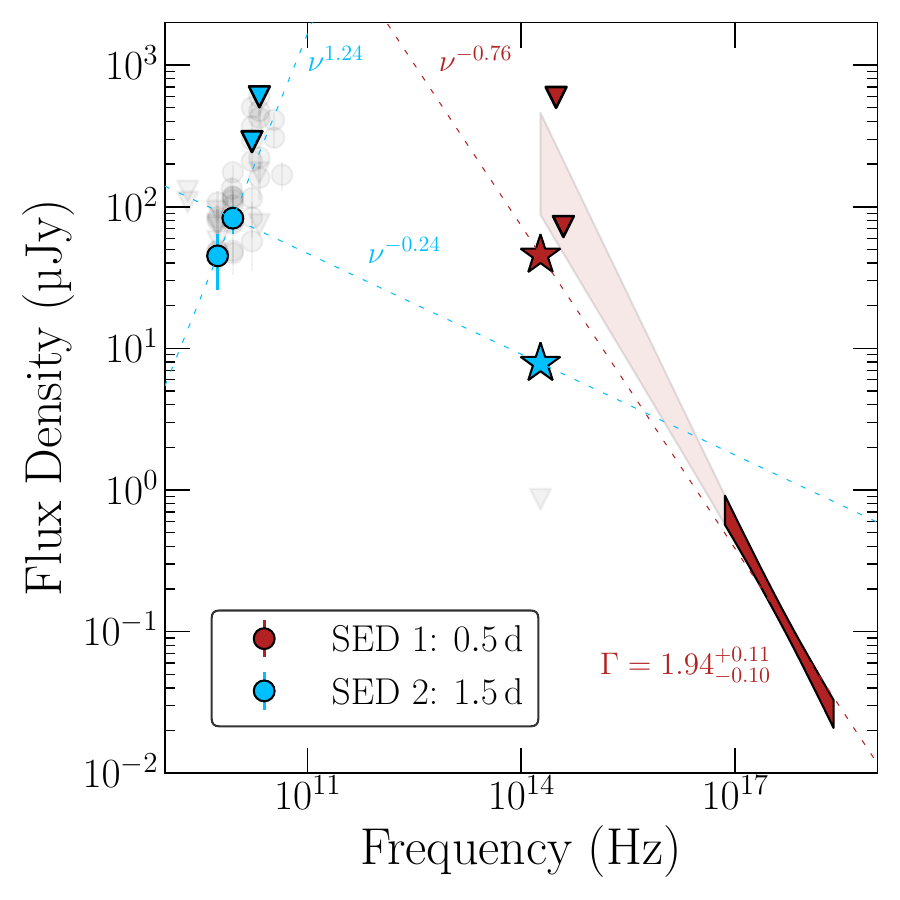}
    \caption{Broadband SED of GRB~230815A, showing two temporal snapshots at approximately 0.5 (red) and 1.5 (blue) days post-burst. 
    The light gray points represent data from later epochs, after 1.5\,days post-burst. 
    Radio detections are shown with circular markers, near-infrared detections are shown with star markers, and all upper limits are shown with inverted triangular markers. 
    The X-ray spectrum (dark red filled) is extrapolated to the near-infrared part of the spectrum (light red filled) using a photon index of $\Gamma = 1.94^{+0.11}_{-0.10}$, where the shaded region corresponds to a 68\% confidence interval. 
    The slopes between various data points in each spectral snapshot are annotated next to the corresponding dashed guide lines. 
    }
    \label{fig:bbsed}
\end{figure}

Considering only the photons detected by \textit{Swift}/XRT while in PC mode, the overall time-averaged photon index was $\Gamma (t_0 + 134\ \text{to}\ 45\,167\,\text{s}) = 1.94^{+0.11}_{-0.10}$, where the mean photon arrival time was $t_0 + 3\,322$\,s. 
When considering the spectrum from the photons that were detected after the apparent X-ray light-curve break, the late-time time-averaged photon index was $\Gamma (t_0 + 6\,081\ \text{to}\ 45\,167\,\text{s}) = 1.6^{+0.4}_{-0.3}$, where the mean photon arrival time was $t_0 + 21\,200$\,s. 
Given the low photon count in this late-time segment, the uncertainties on the photon index are large, preventing us from determining whether the spectrum evolved after the apparent break. 

In Figure~\ref{fig:bbsed} we plot the broadband spectral energy distribution (SED) of the GRB in two early snapshots at 0.5 and 1.5 days post-burst, where there is coverage across multiple windows of the broadband spectrum. 
The darker red shaded region shows the X-ray spectrum of the afterglow across the energy range of the \textit{Swift}/XRT detector, using the measured unabsorbed flux from X-ray photons arriving approximately $0.5$ days post-burst and assuming the aforementioned overall time-averaged photon index $\Gamma = 1.94^{+0.11}_{-0.10}$ (i.e., assuming minimal evolution in the X-ray spectrum up until 0.5 days post-burst). 
In the lighter red shaded region, we extrapolated the X-ray spectrum beyond the detector energy range into the near-infrared part of the spectrum, where we were able to compare against the HAWK-I detection made at the same epoch. 
The near-infrared data point lies outside the shaded region of the extrapolated spectrum (corresponding to a 68\% confidence interval), suggesting the possibility of a break between the X-ray and near-infrared bands at 0.5 days post-burst. 
Examining now the epoch at 1.5 days post-burst (corresponding to SED~2, coloured blue in Figure~\ref{fig:bbsed}), the peak of the spectrum clearly lies between the radio and near-infrared wavelengths. 
The decay of the near-infrared spectrum from 0.5 to 1.5 days post-burst is also not as steep as the $t^{-2.4 \pm 0.2}$ observed in the X-ray; instead, as shown more clearly in Figure~\ref{fig:mwlc} the decay follows a shallower $t^{-1.6 \pm 0.1}$. 

\subsection{Chromatic turnover in the radio light curves}\label{ssec:radiolc}

\begin{table*}
\fontsize{10pt}{12pt}\selectfont
\caption{The nested sampling best-fit parameters for two models describing the multi-frequency radio light curves for GRB~230815A at 5.5, 9, 16.7, and 21.2\,GHz. 
The first model assumes the rise and decay slopes ($\delta_1$ and $\delta_2$) are independent for each frequency, while the second model fixes the rise and decay slopes to be the same for all frequencies. 
Column 1 shows the frequency, columns 2 through 5 show the best-fit parameters and their uncertainties for the model fits (i.e., the median and the 16$^{\text{th}}$/84$^{\text{th}}$ percentile of the posterior distribution), column 6 shows the log evidence of the models, and column 7 shows the log Bayes factor comparing the two models. 
}
\label{tab:radiolc_fits}
\begin{tabular}{|l|cccc|c|c|}
\hline\hline
\multicolumn{7}{|c|}{\textbf{Model 1: Independent $\delta_1$, $\delta_2$ per frequency}} \\
\hline
 & $S_{\rm peak}$ (\textmu Jy) 
 & $t_{\rm b}$ (days) 
 & $\delta_1$ & $\delta_2$ & $\ln \mathcal{Z}$ & $\ln \mathcal{B}_{\rm model1,\,model2}$ \\
\hline
5.5\,GHz & $72.19^{+13.00}_{-9.68}$ & $246.45^{+114.74}_{-156.43}$ & $0.18^{+0.11}_{-0.07}$ & $-0.37^{+0.27}_{-0.52}$ & \multirow{4}{*}{$-173.67$} & \multirow{4}{*}{$7.75$} \\
9\,GHz & $134.90^{+17.87}_{-16.77}$ & $119.68^{+47.59}_{-36.13}$ & $0.22^{+0.08}_{-0.07}$ & $-1.09^{+0.28}_{-0.40}$ & & \\
16.7\,GHz  & $440.10^{+33.25}_{-32.70}$ & $40.80^{+4.85}_{-4.50}$ & $0.70^{+0.15}_{-0.13}$ & $-1.23^{+0.16}_{-0.19}$ & & \\
21.2\,GHz & $580.30^{+48.43}_{-48.68}$ & $42.76^{+5.97}_{-6.13}$ & $0.82^{+0.21}_{-0.18}$ & $-1.42^{+0.33}_{-0.33}$ & & \\
\hline
\multicolumn{7}{|c|}{\textbf{Model 2: Common $\delta_1$, $\delta_2$ across different frequencies}} \\
\hline
 & $S_{\rm peak}$ (\textmu Jy) 
 & $t_{\rm b}$ (days) 
 & $\delta_1$ & $\delta_2$ & $\ln \mathcal{Z}$ & $\ln \mathcal{B}_{\rm model1,\,model2}$ \\
\hline
5.5\,GHz & $105.05^{+32.17}_{-26.13}$ & $134.52^{+181.47}_{-44.46}$ & \multirow{4}{*}{$0.39^{+0.08}_{-0.08}$} & \multirow{4}{*}{$-1.14^{+0.13}_{-0.15}$} & \multirow{4}{*}{$-181.42$} & \multirow{4}{*}{$7.75$} \\
9\,GHz & $171.11^{+16.05}_{-16.28}$ & $93.71^{+20.16}_{-15.98}$ & & & & \\
16.7\,GHz & $368.18^{+25.49}_{-25.15}$ & $47.83^{+6.60}_{-5.56}$ & & & & \\
21.2\,GHz & $468.27^{+36.82}_{-34.61}$ & $51.63^{+7.75}_{-6.40}$ & & & & \\
\hline
\end{tabular}
\end{table*}

The radio light curves at 5.5, 9, 16.7, and 21.2\,GHz in Figure~\ref{fig:mwlc} show that they are evolving as smoothly broken power laws (as expected for a synchrotron afterglow), with indications of a chromatic break that occurs later for lower frequencies. 
Motivated by this, we fit a smoothly broken power-law function, given by Equation~\ref{eq:sbpl} (replacing $S_{\rm break}$ with $S_{\rm peak}$), to the light curves at each frequency, while setting the smoothness parameter to $s=3$ to reduce the degrees of freedom in the model (given the sparsity of the data per frequency). 
Again, following the same nested sampling procedure as described in Section~\ref{ssec:xraybreak}, we fit all the light curves at once under two models with different assumptions. 
In the first model, we let the rise slope $\delta_1$ and the decay slope $\delta_2$ be independently sampled for each frequency. 
In the second model, motivated by the idea that the rise and decay slopes should be similar across all frequencies in the case where the evolution is caused by the passage of a characteristic frequency across the radio band, we fix the rise slope $\delta_1$ and decay slope $\delta_2$ to be common across different frequencies. 
We performed the nested sampling again using uniform priors, with 2\,000 live points, and a stopping criterion on the change in estimated evidence $\hat{\mathcal{Z}}$ between iterations of $\Delta \mathrm{ln}(\hat{\mathcal{Z}}) = 0.05$. 

The resulting posterior distributions of the parameters in the fits of the radio light curves to both models are reported in Table~\ref{tab:radiolc_fits} and are also shown in Figures~\ref{fig:corner_radiolc_sbpl_joint} and \ref{fig:corner_radiolc_sbpl_fixd1d2} as corner plots. 
These show that in both models the break time and peak flux density for the light curve at 5.5\,GHz are poorly constrained, suggesting that the light curve at this frequency may not have or has just recently turned over. 
Both models suggest the presence of a chromatic turnover in the radio light curves: at higher frequencies, the turnover occurs sooner with a higher peak flux density, which is consistent with what may be expected 
from a light-curve evolution that is due to a spectral break crossing the radio band as it moves toward lower frequencies \citep[note that this is applicable in certain configurations, such as when $\nu_\textrm{m} > \nu_\textrm{a}$ or in a stratified CBM; e.g.,][]{Gao2013}. 

The logarithm of the Bayes factor comparing Model 1 (independent $\delta_1$ and $\delta_2$ per frequency) to Model 2 (fixed $\delta_1$ and $\delta_2$ across all frequencies) $\ln \mathcal{B}_{\rm model1, model2} = \ln \mathcal{Z}_{\rm model1} - \ln \mathcal{Z}_{\rm model2} = 7.75$, suggests that Model 1 is better supported by the data than Model 2. 
The key difference between the results from Models 1 and 2 is that both the rise and decay slopes are steeper for higher frequencies when they are allowed to vary independently per frequency. 
As Model 1 is better supported, these results indicate that the chromatic turnover cannot be attributed to just the passage of one characteristic frequency through the observing bands, but instead point to the possibility of two characteristic frequencies -- namely $\nu_\textrm{a}$ and $\nu_\textrm{m}$ -- passing through the observing bands, with their ordering flipping at some point. 
The fit obtained for Model 1 is plotted in the radio light curves in Figure~\ref{fig:mwlc}, along with 50 randomly selected samples from the posterior parameter distribution (indicative of the fit uncertainty). 

\begin{figure*}
    \centering
    \includegraphics[width=0.94\textwidth]{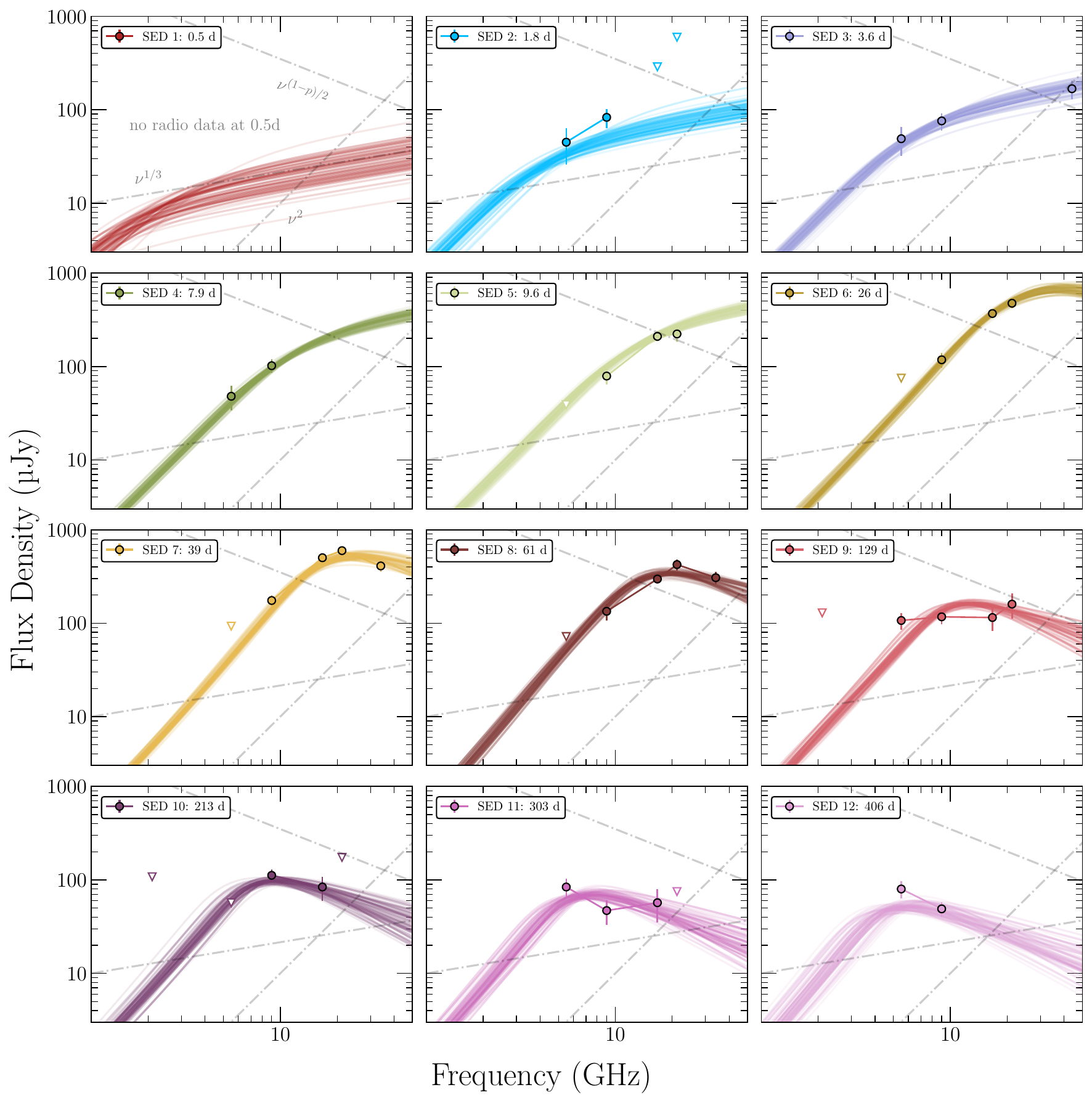}
    \caption{Radio SEDs of GRB~230815A at the 12 different epochs constructed from the temporal windows shown in Figure~\ref{fig:mwlc}.  
    The epoch is labelled in the legend for each panel. 
    The dotted-dashed lines are guide lines showing different spectral slopes. 
    These spectral slopes are labelled in SED 1 (0.5 days post-burst), with $p=2.2$ adopted here (see the text). 
    Detections are shown with filled circular markers, while $3\sigma$ upper limits for non-detections are shown using inverted triangular markers. 
    The thin lines in each panel show model spectra from 100 random posterior samples from the nested sampling procedure performed to constrain our evolving synchrotron model in Subsection \ref{ssec:radiosed_new}. 
    }
    \label{fig:radioseds}
\end{figure*}

\subsection{Radio chromatic evolution as described within the synchrotron paradigm}\label{ssec:radiosed_new}

While our tests in the previous section show that the observed radio evolution is not well described by the passage of a single synchrotron break frequency through the bands, the spectral evolution (see Figure~\ref{fig:radioseds}) clearly shows a transition from a regime where the lowest frequencies are self-absorbed (as suggested by the steeply rising spectral slope visible up to $\sim$60 days post-burst) to one where all frequencies lie on an optically thin branch of the spectrum. 
We therefore propose a more general scenario to describe the system, where two synchrotron frequencies -- the injection frequency $\nu_\mathrm{m}$ and the self-absorption frequency $\nu_\mathrm{a}$ -- cross the radio bands over the observed period, possibly also crossing each other. 
In order to obtain a description of the spectra and light curves that is always smooth, we opted to describe the optically thin synchrotron spectrum as a smoothly broken power law: 
\begin{equation}
    S_{\nu,\rm thin} = S_{\nu_\mathrm{m}}\left[\left(\frac{\nu}{\nu_\mathrm{m}}\right)^{-s_1/3}+\left(\frac{\nu}{\nu_\mathrm{m}}\right)^{s_1(p-1)/2}\right]^{-1/s_1},
\end{equation}
where $S_{\nu_\mathrm{m}}$ is the flux density at the extrapolated meeting point of the two power-law branches \citep{Granot2002}, and $s_1$ is a smoothing parameter; and the frequency-dependent synchrotron self-absorption optical depth as \citep[see, e.g.,][]{Panaitescu2000}:
\begin{equation}
    \tau_{\nu,\rm SSA} = \tau_{\nu_\mathrm{m}}\left[\left(\frac{\nu}{\nu_\mathrm{m}}\right)^{s_2 5/3}+\left(\frac{\nu}{\nu_\mathrm{m}}\right)^{s_2(p+4)/2}\right]^{-1/s_2},
\end{equation}
where $\tau_{\nu_\mathrm{m}}$ is the optical depth at $\nu_\mathrm{m}$, and $s_2$ is another smoothing parameter. From basic radiative transfer (assuming the emitting and absorbing electron populations to be the same), the resulting spectrum is then:
\begin{equation}
    S_{\nu} = S_{\nu,\mathrm{thin}}\frac{1-\exp(-\tau_{\nu,\rm SSA})}{\tau_{\nu,\rm SSA}}.
\end{equation}
The benefit of this description over the more widely adopted \citet{Granot2002} one is that it does not suffer from a discontinuous behavior when $\nu_\mathrm{m}$ and $\nu_\mathrm{a}$ cross each other.

\begin{table*}
    \caption{Nested sampling results for the evolving self-absorbed synchrotron spectrum model of Subsection \ref{ssec:radiosed_new}. Below each parameter, we report the support of the prior (uniform for all parameters in the reported range) and a summary of the result, which consists of the median and 68\% symmetric credible interval of the marginalised posterior probability of that parameter. The full corner plot is shown in Figure \ref{fig:corner_semi_emp}.}
    \centering\footnotesize
    \begin{tabular}{|r|ccccccccc|}
    \hline\hline
       Parameter & $s_1$ & $s_2$ & $\log(S_{\nu_\mathrm{m},0}/\mathrm{\textrm{\textmu} Jy})$ & $\log(\tau_{\nu_\mathrm{m},0})$ & $\log(\nu_{\mathrm{m},0}/\mathrm{GHz})$ & $\alpha_\mathrm{S}$ & $\alpha_\mathrm{\tau}$ & $\alpha_\mathrm{m}$ & $p$  \\
       \hline
       Prior range & $[0.5,3]$ & $[0.5,3]$ & $[1,6]$ & $[-3,6]$ & $[-3,3]$ & $[-4,0.5]$ & $[0,6]$ & $[-4,4]$ & $[2.01,4]$\\
       \hline
       Result & $2.63_{-0.43}^{+0.26}$ & $0.58_{-0.06}^{+0.10}$ & $3.10_{-0.09}^{+0.10}$ & $2.67_{-0.37}^{+0.37}$ & $0.32_{-0.17}^{+0.16}$ & $0.19_{-0.12}^{+0.13}$ & $4.71_{-0.51}^{+0.56}$ & $-2.12_{-0.29}^{+0.26}$ & $2.57_{-0.33}^{+0.41}$\\
       \hline
    \end{tabular}    
    \label{tab:semi_emp_fit}
\end{table*}

For what concerns the time evolution of the spectral parameters $S_{\nu_\mathrm{m}}$, $\tau_{\nu_\mathrm{m}}$ and $\nu_\mathrm{m}$, we opted for a simple description in terms of power laws, namely:
\begin{eqnarray}
& S_{\nu_\mathrm{m}}(t) = S_{\nu_\mathrm{m},0}\left(\frac{t}{t_0}\right)^{\alpha_\mathrm{S}},\\
& \tau_{\nu_\mathrm{m}}(t) = \tau_{\nu_\mathrm{m},0}\left(\frac{t}{t_0}\right)^{\alpha_{\tau}},\\
& \nu_\mathrm{m}(t) = \nu_{\mathrm{m},0}\left(\frac{t}{t_0}\right)^{\alpha_\mathrm{m}},
\end{eqnarray}
where we chose the reference time to be $t_0=100$ days. The resulting model has nine parameters -- namely $s_1$, $s_2$, $S_{\nu_\mathrm{m},0}$, $\tau_{\nu_\mathrm{m},0}$, $\nu_{\mathrm{m},0}$, $\alpha_\mathrm{S}$, $\alpha_\mathrm{\tau}$, $\alpha_\mathrm{m}$, and $p$. We estimated the posterior probability distributions of these parameters adopting the same nested sampling technique as in previous sections, assuming relatively wide and uninformative priors, as detailed in Table \ref{tab:semi_emp_fit}, where we also report summaries of the fit results. The full corner plot is shown in Figure \ref{fig:corner_semi_emp} in the Appendix. 
The thin lines in each panel of Figure \ref{fig:radioseds} show model spectra from 100 random posterior samples, demonstrating that the model successfully reproduces the observed evolution. 

\begin{figure}
    \centering
    \includegraphics[width=\columnwidth]{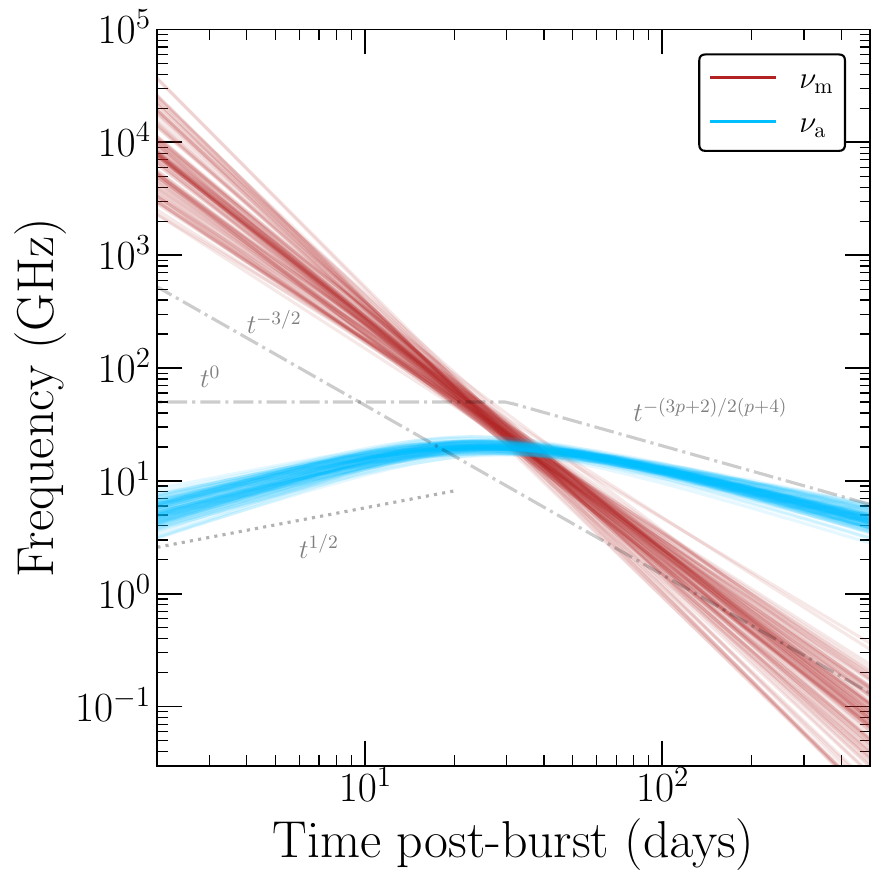}
    \caption{Time evolution of the injection ($\nu_\mathrm{m}$; red lines) and self-absorption ($\nu_\mathrm{a}$; blue lines) frequencies from 100 random posterior samples of the evolving synchrotron spectrum model of Subsection \ref{ssec:radiosed_new}. The gray dotted-dashed lines show the expected evolutions of the two frequencies in the self-similar phase of a \citet{Blandford1976} blast wave (valid for a homogenous ISM environment), as given in \citet{Granot2002}, assuming $p=2.6$. The dotted line shows $t^{1/2}$, which approximates the initial evolution of $\nu_\mathrm{a}$.}
    \label{fig:num_nua}
\end{figure}

The temporal evolution of the break frequencies $\nu_\mathrm{m}$ and $\nu_\mathrm{a}$ (defined as the frequency for which $\tau_{\nu,\mathrm{SSA}}=1$) implied by the model is shown in Figure \ref{fig:num_nua}. The dotted-dashed gray lines in the figure show the expected evolution for a \citet{Blandford1976} blast wave with constant microphysical parameters \citep[][where the $\nu_\mathrm{m}$ evolution is valid for both the ISM and wind, while that of $\nu_\mathrm{a}$ is shown only for the ISM case here]{Granot2002}, where $t^0$ (respectively, $t^{-(3p+2)/2(p+4)}$) refers to the evolution of $\nu_\mathrm{a}$ prior to (respectively, after) being crossed by $\nu_\mathrm{m}$. 
The $\nu_\mathrm{m}$ evolution preferred by the model is somewhat steeper than the expected one ($\alpha_\mathrm{m}\sim -2.1\pm0.3$ instead of the expected $-1.5$); the evolution of $\nu_\mathrm{a}$ is slightly increasing before the $\nu_\mathrm{m}$ crossing (approximately as $t^{1/2}$, instead of the expected $t^0$), and it is somewhat shallower than the expected one after the crossing. In addition, the result shows a mild preference for a slightly increasing $S_{\nu_\mathrm{m}}$, quantitatively $\alpha_\mathrm{S}\sim 0.2\pm 0.13$, in contrast with the expected constant (ISM) or slowly decreasing (wind) behaviour \citep{Granot2002}. In general, the non-standard evolution of the break frequencies could reflect a deviation of the shock dynamics from the standard self-similar expansion (e.g., energy injection or sideways expansion) or an evolution of the microphysical parameters. We note that a very similar result is found in an analogous analysis of late-time radio observations of GRB~241025A (Giarratana et al., \textit{in preparation}), possibly pointing to a widespread behaviour among GRB radio afterglows.

\section{Discussion}\label{sec:discussion}
\subsection{A Unified Physical Picture of GRB~230815A}\label{ssec:unifiedpicture}

The X-ray break (Subsection~\ref{ssec:xraybreak}) at $10.9^{+3.2}_{-2.6}$\,ks (or $0.13^{+0.04}_{-0.03}$ days) post-burst transitions from a pre-break temporal slope of $\delta_1 = -1.26^{+0.05}_{-0.04}$ to a post-break temporal slope of $\delta_2 = -2.44^{+0.16}_{-0.19}$. 
This change in temporal slope after the break $\Delta \delta = \vert\delta_2 - \delta_1\vert = 1.18^{+0.24}_{-0.20}$ cannot be explained by the passing of the cooling break $\nu_{\rm c}$ through the X-ray band, since this is expected to yield $\Delta \delta = 1/4$ in both the homogeneous and stellar wind scenarios \citep[e.g.,][]{Granot2002}. 
To strengthen this point, if the break were due to the passing of $\nu_{\rm c}$, for any given value of the electron spectral index $p$ (defined as the power-law index of the synchrotron electron energy distribution $\frac{dN}{dE} \propto E^{-p}$), the steepest resulting temporal slope should go as $\propto t^{\frac{1-3p}{4}}$, which would still require a very extreme $p \approx 3.5$ \citep[e.g.,][]{Wang2015}, inconsistent with the values inferred from the photon index (discussed below). 
Instead, we propose that the break observed in the X-ray is an early jet break. 
Analytical solutions for the jet break, considering the edge effect and lateral spreading, suggest the slopes of the temporal decay in both the homogeneous and stellar wind environment can be as steep as $\propto t^{-p}$ \citep[in the $\nu > \nu_\textrm{m}$ and $\nu > \nu_\textrm{c} > \nu_\textrm{m}$ spectral regimes;][]{Rhoads1999,Sari1999,Livio2000}; however, when considering only the edge effect with no lateral spreading, there will be a more shallow change of 3/4 (respectively, 1/2) in the temporal index with respect to the pre-break slope in the homogeneous (respectively, wind) environment \citep[e.g.,][]{Gao2013}. 
A more detailed numerical study of the jet break in the homogeneous density CBM environment yields a post-jet-break decay index of $0.28-1.29p$ for an on-axis observer of a jet with typical opening angle of $0.1$\,rad in the $\nu > \nu_\textrm{c}$ spectral regime \citep{vanEerten2013}; this constrains the electron spectral index to $p = 2.11^{+0.15}_{-0.12}$. 
While a similar numerical study comprehensively quantifying the pre- and post-jet break behaviour is not available for a stratified CBM environment, other studies have indicated that the post-break decay slope in a stellar wind environment is expected to be similar or shallower by up to a factor of ${\sim} 2/3$ \citep[e.g., see figure 10 in][]{DeColle2012}, allowing for a steeper yet physical $p \sim 2.6$.

We then consider the measured X-ray photon index (see Subsection~\ref{ssec:bbsed}) of $\Gamma = 1.94^{+0.11}_{-0.10}$, which implies a spectral index of $\beta_{\rm x} = -0.94^{+0.10}_{-0.11}$. 
We can obtain an estimate of $p$ using standard closure relations \citep[e.g.,][]{Sari1998} by considering the possibilities of whether the X-ray band $\nu_{\rm x}$ is located below or above $\nu_{\rm c}$. 
In the scenario that $\nu_{\rm x} < \nu_{\rm c}$, the spectral index goes as $S_\nu \propto \nu^{\frac{1-p}{2}}$, so the spectral index $\beta_{\rm x}$ implies $p = 2.88^{+0.22}_{-0.20}$, which is inconsistent with the lower $p$ estimated from the post-jet-break temporal index in a homogeneous environment but possibly consistent with the steeper estimate in the stellar wind environment (quoted above). 
However, the expected pre-jet-break temporal index in a stellar wind environment goes as $\propto t^{\frac{1-3p}{4}}$ so the $p = 2.88^{+0.22}_{-0.20}$ inferred from the spectral index would imply a temporal index of $-1.91^{+0.15}_{-0.17}$, which is too steep to explain the much shallower pre-jet-break decay of $-1.26^{+0.05}_{-0.04}$. 
Therefore, it is unlikely that $\nu_{\rm x} < \nu_{\rm c}$ at 0.5 days post-burst; indeed, in Figure~\ref{fig:bbsed} the X-ray spectrum extrapolated to the near-infrared suggests that it is possible (at >$1\sigma$ confidence) that the near-infrared and X-ray data points do not lie on the same spectral segment.
In the scenario $\nu_{\rm c} < \nu_{\rm x}$, the spectral index goes as $S_\nu \propto \nu^{-\frac{p}{2}}$, implying a lower estimate for $p = 1.88^{+0.22}_{-0.20}$, while the pre-jet-break temporal indices go as $\propto t^{\frac{2-3p}{4}}$ in both the homogeneous and stellar wind CBM environment, yielding an estimate of $p = 2.35^{+0.05}_{-0.07}$, consistent at the 90\% confidence level with the estimate from the spectral index. 
We note that the tendency toward lower values of $p$ inferred here gives preference to the homogeneous environment over the stellar wind environment. 
Using this information and adopting $p = 2.2$, taken as the middle point between the two estimates of $p$ from the X-ray temporal and spectral indices, respectively, and that $\nu_{\rm o} < \nu_{\rm c} < \nu_{\rm x}$, we estimate the cooling break at the epoch of ${\sim} 0.5$ days to be 
${\sim} 10^{16}$\,Hz, with the flux density at the break to be ${\sim}1\,$\textmu Jy. 

After the jet break, the decay of the near-infrared slope as $t^{-1.6\pm0.1}$ (see Subsection~\ref{ssec:bbsed}) is shallower than the post-break slope observed in the X-ray decaying as $t^{-2.4\pm0.2}$. 
If the X-ray and near-infrared bands are observing radiation from the same emitting component, their post-jet-break decay slopes should match, since the predicted decay slopes should be similar in both the spectral regimes of $ \nu_\textrm{c} > \nu > \nu_\textrm{m}$ and $ \nu > \nu_\textrm{c} > \nu_\textrm{m}$ for both the analytical \citep[e.g.,][]{Sari1999} and numerical \citep[e.g.,][]{vanEerten2013} predictions that consider lateral spreading. 
Without considering lateral spreading (and just the edge effect), the difference in the temporal decay in the $ \nu_\textrm{c} > \nu > \nu_\textrm{m}$ and $ \nu > \nu_\textrm{c} > \nu_\textrm{m}$ spectral segments should differ by $1/4$ \citep[e.g.,][]{Gao2013}, which is insufficient to explain the difference of $0.8 \pm 0.2$ observed between the post-jet-break X-ray and near-infrared decay slopes. 
The presence of a flatter slope at the near-infrared wavelengths could indicate an additional component; due to the limited spectral and temporal coverage around this time and wavelength, it is difficult to constrain the properties and origins of this excess -- e.g., whether it may be related to the origins of the radio emission (discussed below) or something else. 

We next consider the radio light curves (Subsection~\ref{ssec:radiolc}) and the evolving SEDs (Subsection~\ref{ssec:radiosed_new}). 
The chromatic turnover in the radio light curves is indicative of the passing of a (or multiple) characteristic synchrotron frequency (frequencies) through the observing band. 
The lack of any observed achromatic breaks in the radio light curves suggests that even at 400 days post-burst, there is no evidence for changes to a new dynamical regime, such as a transition into the non-relativistic regime or into a different surrounding density profile. 
Most surprisingly, the decays of the radio light curves (which are all $\gtrapprox -1.42$) are much shallower than that observed in the X-ray following the jet break; if the X-ray and radio afterglow emission share the same origins, their post-jet-break decay slopes (after the spectral peak $\textrm{max}(\nu_{\rm m},\nu_{\rm a})$ passes through the band) should be the same. 
Furthermore, in the post-jet-break scenario considering both the edge effect and lateral spreading, the light curves are expected to decay according to power laws in all synchrotron spectral segments \citep[for both the homogeneous and stellar wind scenarios, as $t^{-p}$ at $\nu > \nu_{\rm m}$ and $\nu > \nu_\textrm{c} > \nu_\textrm{m}$, as $t^{-1/3}$ at $\nu_{\rm a} < \nu < \nu_{\rm m}$, and as $t^{0}$ at $\nu < \nu_{\rm a}$; e.g.,][]{Sari1999,Livio2000}, which is in contrast to the rising light curves observed at each radio frequency. 
However, if lateral spreading is not considered to be significant until much later and only the edge effect is considered \citep[e.g., see][and references therein]{Gao2013}, there is only one scenario (in the $\nu_\mathrm{a} < \nu_\mathrm{m}$ configuration that is applicable to our rising radio light curves; see Figure~\ref{fig:num_nua}) where an increasing light curve is possible -- i.e., in the spectral regime of $\nu < \nu_\mathrm{a}$ in a stellar wind environment, the light curve is predicted to go as $t^{1/2}$. 
Even then, our data are incompatible with this picture for the post-jet-break radio behaviour in three ways: 
(i) the 21.2\,GHz light curve should be in the $\nu_\mathrm{a} < \nu < \nu_\mathrm{m}$ spectral segment, which should be decreasing as $t^{-1/2}$, unlike the $\nu < \nu_\mathrm{a}$ segment; (ii) after $\mathrm{max}(\nu_\mathrm{a},\nu_\mathrm{m})$ passes the observing band, the light curve should decrease according to $t^{-(3p+1)/4 \approx -2}$, compared to the $t^{-1.42}$ (and shallower) we observe in our light curves after their turnovers; and (iii) it is far more likely (as discussed in the subsequent paragraphs) that the GRB exploded in a homogeneous ISM and not a wind environment.
The observed radio behaviour is therefore difficult to reconcile with the jet break observed in the X-ray data. 

This suggests the radio emission comes from a different emission origin to that of the X-rays. 
We summarise the key features of the observed radio emission as follows: 
(1) chromatic turnover in the radio light curves with an earlier turnover and higher peak flux density at higher frequencies; 
(2) temporal rise and decay indices for the radio light curves of approximately $0.4^{+0.1}_{-0.1}$ and $-1.1^{+0.1}_{-0.2}$, respectively, with both the rise and decay being sharper at higher frequencies, suggesting the passage and crossing of both $\nu_\textrm{a}$ and $\nu_\textrm{m}$ across the observing band; 
(3) the radio spectrum is initially self-absorbed and then evolves to an optically thin slope of $(1-p)/2$, where $p = 2.57^{+0.41}_{-0.33}$, with the passage of both $\nu_\textrm{a}$ and $\nu_\textrm{m}$ through the observing band and $\nu_\textrm{a}$ crossing $\nu_\textrm{m}$ at approximately ${\sim} 20$\,GHz and  ${\sim} 50$\,days post-burst; 
(4) $\nu_\mathrm{a}$ evolves approximately as $\nu_\mathrm{a} \propto t^{1/2}$ prior to crossing $\nu_\mathrm{m}$ (this is closer to the $\nu_\mathrm{a} \propto t^0$ expected in the homogeneous scenario than the $\nu_\mathrm{a} \propto t^{-3/5}$ expected in the wind scenario) and shallower than the temporal evolution of $t^{-(3p+2)/2(p+4)}$ expected for a \citet{Blandford1976} blast wave with constant microphysical parameters after crossing $\nu_\mathrm{m}$ in both the homogeneous and wind scenarios \citep[e.g.,][]{Granot2002}; 
and (5) $\nu_\textrm{m}$ evolves as $\nu_\textrm{m} \propto t^{-2.1\pm 0.3}$, which is steeper than the decay of $\nu_\textrm{m} \propto t^{-1.5}$ expected in both the homogeneous and wind scenarios, while $S_{\nu_\mathrm{m}}$ evolves as $S_{\nu_\mathrm{m}} \propto t^{0.2\pm 0.13}$, which is slightly increasing and is more consistent with the expected constant ISM behaviour compared with the slowly decreasing $S_{\nu_\mathrm{m}} \propto t^{-1/2}$ expected in the wind environment \citep[e.g.,][]{Granot2002}.

As previously mentioned, these radio afterglow properties display minor deviations away from the standard shock dynamics expected from the standard self-similar expansion \citep[e.g.,][]{Granot2002}, which could be indicative of additional energy injection, a multi-component jet structure, sideways expansion, or an evolution of the microphysical parameters. 
Nevertheless, as shown in the aforementioned points (4) and (5), the properties are more consistent with expanding into a homogeneous density environment than a stellar wind environment. 
We note here that while the $p=2.57^{+0.41}_{-0.33}$ in point (3) is slightly higher than the $p \approx 2.2$ determined earlier from the X-ray data, they are broadly consistent with each other, since the  uncertainties on $p$ from the evolving radio synchrotron model in Subsection~\ref{ssec:radiosed_new} are larger, given that the optically thin spectrum is relatively unconstrained by the data; we further note that the optically thin SEDs 10 and 11 in Figure~\ref{fig:radioseds} have slopes of approximately $-0.6$, which is consistent with $p \approx 2.2$. 
Critically, we infer from these radio light curves that the blast-wave evolution is inconsistent with the post-jet-break scenario, which is in direct conflict with the jet break observed in the X-ray light curve, because: 
(i) the radio decay slopes are much shallower than the post-jet-break decay slope in the X-ray light curve when they should be similar post jet-break even if they are in different spectral segments; 
(ii) the observed rising radio light curve is not consistent with what is expected in the post-jet-break scenario (see the prior discussion about the expected light curve behaviour in the post-jet-break scenario when considering both the scenario with and without lateral spreading); and 
(iii) the expected decay of $S_{\nu_{\rm m}} \propto t^{-1}$ in the post-jet-break scenario is inconsistent with the slightly positive increase we see in $S_{\nu_{\rm m}}$ as a function of time \citep{Sari1999,Livio2000,vanEerten2013,Gao2013}. 

There are two options to reconcile this: the jet break at radio frequencies is delayed with respect to the X-ray band, or the structure of the jet is different to the standard top-hat model. 
In the former scenario, it is possible that the jet break at radio frequencies could be delayed with respect to the X-ray band, simply because the radio-emitting electrons are located in regions of the outflow farther away from the jet edge (which is tracked better by the high-energy X-ray electrons) due to limb-brightening and self-absorption effects \citep{vanEerten2011}. 
However, these effects would typically delay the jet break by a factor of a few or up to an order of magnitude \citep[e.g.,][]{vanEerten2011,DeColle2012}, which is insufficient to explain the greater than 2 orders of magnitude delay in the jet break time we see between the X-ray and radio light curves for GRB~230815A, thus disfavouring this scenario.  
We propose instead that the remaining option for reconciling this involves a two-component jet \citep[e.g.,][]{Berger2003,Starling2005,Racusin2008,vanderHorst2014}: a fast, narrow component responsible for the early X-ray jet break, and a slower, wide component responsible for the delayed (or lack of) jet break observed at radio frequencies even at 400 days post-burst. 
Comprehensive modelling of the multi-wavelength light curves using a two-component jet model (where the dynamics of the two jets are coupled together, rather than independent, as assumed in this manuscript) can verify this and characterise the components of the system, but this is beyond the scope of this work. 
We stress that for any future multi-wavelength modelling pursuits, the lack of a redshift measurement and the unconstrained location of the cooling break corresponding to the second wider component will lead to large uncertainties in constraining the outflow physical parameters. 

We note that in order to more clearly highlight the features of the afterglow in this picture, we invoked the two-component jet. 
However, it is more likely that the two-component jet is a simplified approximation for a more realistic structured jet, which has a continuous (rather than with two distinct components) angular stratification in its kinetic energy \citep[see, e.g.,][]{Lamb2021,Salafia2022b}. 

\subsection{Inferred Dynamical and Microphysical Properties}\label{ssec:inferences}

Even without a redshift measurement and standard multi-wavelength afterglow modelling, light curves and broadband spectral snapshots enable us to make some inferences on the dynamical and microphysical properties of the GRB that would otherwise not be possible without a comprehensive high-cadence follow-up of the radio afterglow.  

\subsubsection{Jet Geometry}
\begin{figure}
    \centering
    \includegraphics[width=0.94 \linewidth]{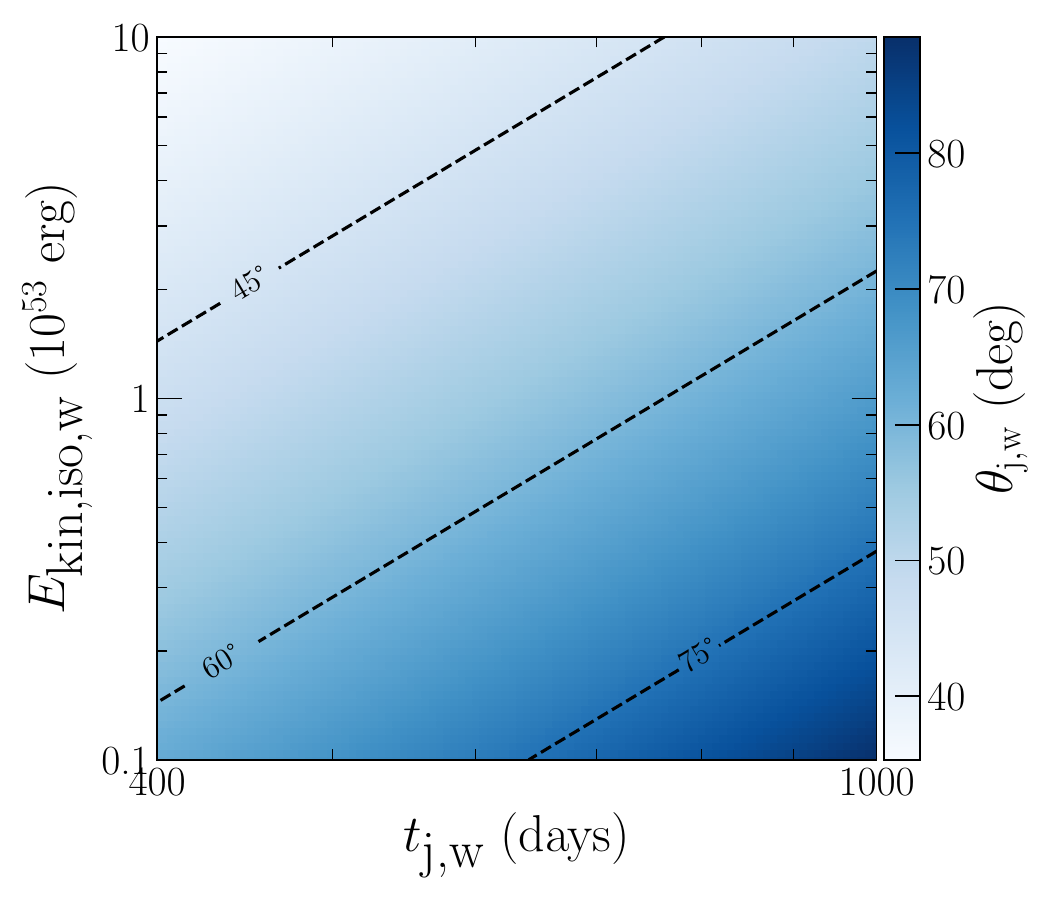}
    \caption{The half-opening angle of the wider component $\theta_{\rm j,w}$\,(in degrees) as a function of the jet break $t_{\rm j,w}$\,(days) and isotropic-equivalent kinetic energy  $E_{\rm k,iso,w}$\,(erg). 
    The dashed lines indicate the combinations of $t_{\rm j,w}$ and $E_{\rm k,iso,w}$ that will yield $\theta_{\rm j,w}$ of $45\degr$, $60\degr$, and $75\degr$. 
    We assume here the redshift and CBM density from Equation~\ref{eq:jetbreak}, effectively performing a relative scaling from the $2.1\degr$ narrow jet. 
    }
    \label{fig:opening_angle_wide}
\end{figure}

Starting with the X-ray jet break, we can constrain the half-opening angle of the narrow jet component \citep[e.g.,][]{Sari1999,Zhang2009}:
\begin{equation}
    \begin{split}
    & \theta_{\rm j} = 2.1\degr \left(\frac{t_{\rm j}}{0.1\ \textrm{days}} \right)^{3/8} \left(\frac{1+z}{1+1.5} \right)^{-3/8} \\
    & \indent\indent \cdot \left(\frac{E_{\rm kin,iso}}{10^{53}\,\textrm{erg}} \right)^{-1/8} \left(\frac{n_{0}}{1\,\textrm{cm}^{-3}} \right)^{1/8},
    \end{split}
    \label{eq:jetbreak}
\end{equation}
where $t_{\rm j}$ is the jet break time in days post-burst in the observer frame, $z$ is the redshift,\footnote{The redshift is assumed here to be $z=1.5$, noting that the median inferred redshift for GRBs is approximately $z\approx1.7$, based on the sample found at \url{https://www.mpe.mpg.de/~jcg/grbgen.html} compiled by Jochen Greiner.}  
$E_{\rm kin,iso}$ is the isotropic-equivalent kinetic energy of the outflow,\footnote{We estimate $E_{\rm kin,iso}$ by assuming a prompt efficiency $\epsilon_\gamma = E_{\gamma,\text{iso}}/(E_{\gamma,\text{iso}}+E_{\text{kin,iso}} + E_{\text{kin,iso,w}}) = 15\%$ \citep[e.g.,][]{Beniamini2016}. Here, $E_{\gamma,\text{iso}} \approx \frac{4\pi d_\textrm{L}^2}{(1+z)} S_\gamma$, where $d_\textrm{L}$ is the luminosity distance assuming $z=1.5$, $S_\gamma = 1.28\pm0.01$\,erg\,cm$^{-2}$ is the fluence detected by \textit{Fermi}/GBM across the energy band $10 - 1\,000$\,keV \citep{MailyanGCN34440}, and $E_{\text{kin,iso,w}}$ is the isotropic-equivalent kinetic energy from the wider component. 
This yields $E_{\rm kin,iso}+E_{\rm kin,iso,w} \approx 4\times 10^{53}$\,erg, so we estimate within an order of magnitude $E_{\rm kin,iso} = 10^{53}$\,erg.} and $n_0$ is the density of the CBM environment. 
The early jet break implies a very narrow jet opening angle, consistent with the narrow end of the long-GRB distribution \citep{Ryan2015,Wang2018} but inconsistent with the distribution of short-GRB jet opening angles inferred from previous sample studies \citep{Fong2015}. 
If we now consider the wider component responsible for the radio emission in the two-component jet scenario, we can constrain how much wider the jet opening angle is for this component $\theta_{\rm j,w}$ (compared to the narrow component) as a function of the wider component's jet break time $t_{\rm j,w}$, which is at least greater than $400$ days post-burst, and the isotropic-equivalent kinetic energy $E_{\rm kin,iso,w}$, which we varied from $1/10$ to 10 times that of the narrower component $E_{\rm kin,iso}$. 
Precisely how much of the kinetic energy is in the wider component depends then on its formation, which is uncertain (hence why the allowable range for $E_{\rm kin,iso,w}$ was varied over 3 orders of magnitude) -- e.g., if it is interpreted as a cocoon, then it depends on how efficiently the energy can be deposited as the jet head propagates through the vestiges of the stellar envelope \citep{Matzner2003,Bromberg2011,Salafia2022b}. 
Our constraints are shown in Figure~\ref{fig:opening_angle_wide}. 
We find that in all scenarios, the half-opening angle of the wider component should exceed $35\degr$ (assuming the same redshift and CBM density as in Equation~\ref{eq:jetbreak}). 
The ratio comparing the half-opening angle of the wider component to the narrow component is:  
\begin{equation}
    \frac{\theta_{\rm j,w}}{\theta_{\rm j}} \approx 17 \left(\frac{t_{\rm j,w}}{400\ \textrm{days}} \right)^{3/8} \left(\frac{E_{\rm kin,iso,w}/E_{\rm kin,iso}}{10} \right)^{-1/8} ,
    \label{eq:jetratio}
\end{equation}
or, equivalently, $\theta_{\rm j,w} \gtrapprox 17\, \theta_{\rm j}$. 
For comparison, previous GRBs that invoke a two-component jet to explain their afterglow emission have a wide range of narrow- and wide-component half-opening angles: e.g., GRB~030329 has larger half-opening angles of $\theta_{\rm j} = 5\degr$ and $\theta_{\rm j,w} = 17\degr$ \citep{Berger2003}, while GRB~080319B has smaller half-opening angles of $\theta_{\rm j} = 0.2\degr$ and $\theta_{\rm j,w} = 5\degr$ \citep{Racusin2008}. 

\subsubsection{(Lack of) Non-relativistic Transition}
The $400$-day afterglow also reveals no change to a new dynamical regime, such as approaching a different CBM density profile or transitioning into the non-relativistic regime, since there is no evidence for any achromatic breaks in the multi-frequency light curves. 
We first note that the GRB properties are consistent with exploding in a homogeneous environment, inferred from as early as the X-ray data. 
This homogeneous environment could be indicative of: (i) the part of the stellar wind that has been compressed by the reverse shock from the wind's interaction with the ISM; or (ii) a wind-termination-shock radius that is located quite close to the progenitor, such as in the case where the CBM environment is dense or the stellar wind has low momentum \citep{RamirezRuiz2005}. 

The lack of observed achromatic break also means the blast wave has not entered the non-relativistic regime, which often happens at time:
\begin{equation}
t_{\rm NR} \approx 425 \left(\frac{1+z}{1+1.5} \right) \left(\frac{E_{\rm kin,iso,w}}{10^{53}\,\textrm{erg}} \right)^{1/3}\left(\frac{n_{0}}{1\,\textrm{cm}^{-3}} \right)^{-1/3}\,\textrm{days},
\label{eq:nonrel}
\end{equation}
where the time of the non-relativistic transition is given in the observer frame \citep[e.g.,][]{Piran2004,Zhang2009,BD2015}. 
From this equation, a very typical time for the transition would be approximately 400 days post-burst, which is consistent with the lack of a clear transition observed after 400 days post-burst for GRB~230815A (here, we assume in the equation a similar isotropic-equivalent kinetic energy for the narrow and wide jet components, both approximately $10^{53}$\,erg).
This indicates that, where possible, a GRB should be followed for many years post-burst to probe the physical evolution of GRBs in this regime. 
In certain configurations -- e.g., bursts with substantially lower than typical kinetic energies or higher densities -- the non-relativistic transition occurs much earlier, making the follow-up requirements less stringent; for GRB~970508, $t_{\rm NR} \approx 100$\,days \citep{Frail2000}, and for GRB~030329, $t_{\rm NR} \approx 50$\,days \citep{Frail2005,Resmi2005,vanderHorst2005}. 
After the transition to the non-relativistic regime at $t_{\rm NR}$, the blast wave can be described with the Sedov-von Neumann-Taylor self-similar solution and results in a flattening of the light curve \citep{Frail2000,Zhang2009,Sironi2013}. 
The transition is smooth, making it difficult to pinpoint the light-curve break, but after the transition, accurate calorimetric measurements can be made without degeneracy from the jet geometry. 

While we cannot make these calorimetric measurements for GRB~230815A, since it has not clearly entered the non-relativistic regime, we can use Equation~\ref{eq:nonrel} to constrain the ratio between the blast-wave kinetic energy and the CBM density. 
This ratio determines the Sedov length \citep{Blandford1976} -- $l_{\rm S} \equiv \left(\frac{E_{\rm kin,iso,w}}{\frac{4}{3}\pi n_0 m_{\rm p}c^2}\right)^{1/3}$, where $m_{\rm p}$ is the proton mass and $c$ is the speed of light -- which is the fundamental length scale governing the expansion of the blast wave, e.g., from which the radius and expansion speed of the blast wave can be determined \citep[e.g.,][and references therein]{BD2015,Granot2018}.
With $t_{\rm NR} > 400$ days (and Equation~\ref{eq:nonrel}), we constrain the ratio between the blast-wave kinetic energy and the CBM density to be: 
\begin{equation}
    \log[(E_{\rm kin,iso,w}/n_0)/{\rm erg\ cm}^3] > 54.1 - 3\log(1+z). 
\end{equation}
This gives us $\log[(E_{\rm kin,iso,w}/n_0)/{\rm erg\ cm}^3] > 52.9$ if we take $z=1.5$ as before, which disfavours particularly low blast-wave kinetic energies and high CBM densities. 

\subsubsection{Universality in Electron Acceleration Processes}

Even without multi-wavelength modelling of the afterglow due to the aforementioned difficulties, we can still gain valuable insights into the microphysical evolution of the afterglow by tracing the synchrotron peak at radio frequencies \citep{Beniamini2017,Duncan2023}. 
In particular, we assume that the turnover of the light curves at the radio frequencies is due to the passage of $\nu_{\rm m}$ across the observing frequency and obtain measurements of the time and flux density of the radio synchrotron peak at four different frequencies (as shown in Table~\ref{tab:radiolc_fits}). 
With this, we can evaluate the parameter $\Psi$, which relates the observed radio peaks to a set of microphysical parameters, effectively acting as a proxy for $\epsilon_{\rm e}$, the fractional shock energy in the electrons \citep{Beniamini2017,Duncan2023}. 
In the homogeneous ISM environment, this relation is given by:
\begin{align}
\label{eq:psi_ism}
\Psi_{\text{ism}} & = 
\left(
\frac{261.4(1+z)^{1/2} \nu_{\rm p} t_{\rm p}^{3/2} E^{1/2}_{\gamma,\text{iso},53}}{10^{15} d_{28}^{2} S_{\nu_{\rm p}} \max\left(1, t_{\rm p}/t_{\rm j}\right)^{1/2}}  
\right)^{1/2}  \\
& = \frac{p-2}{0.177(p-1)} \left( \frac{p-0.67}{p+0.14} \right)^{1/2}
\left( \frac{1 - \epsilon_\gamma}{\epsilon_\gamma} \right)^{-1/4} \notag \\ & \indent \cdot n_0^{-1/4} \epsilon_{\rm e}\,\xi^{-3/2}_{\rm e},
\end{align}
where $\nu_{\rm p}$ is the peak frequency in Hz,\footnote{The unit for this was mistakenly stated as GHz in \citet{Duncan2023}.} $t_{\rm p}$ is the peak time in days, $S_{\nu_{\rm p}}$ is the peak flux density in mJy, $E_{\gamma,\text{iso}} = 10^{53} E_{\gamma,\text{iso},53}\,\textrm{erg}$ is the isotropic-equivalent energy released from the high-energy prompt emission, $d_{28}$ is the luminosity distance to the GRB in units of $10^{28}$\,cm, $\epsilon_\gamma = E_{\gamma,\text{iso}}/(E_{\gamma,\text{iso}}+E_{\text{kin,iso}})$ is the energy efficiency of the prompt emission, and $\xi_{\rm e}$ is the fraction of electrons that are accelerated by the shock into the synchrotron power-law distribution. 

The derived value for the parameter $\Psi$ can then be used to evaluate the parameter $\chi$ \citep{Duncan2023}, which effectively acts as a proxy for $\gamma_{\rm m}$, the minimum Lorentz factor of the electron spectral distribution (corresponding to the synchrotron injection frequency). 
In the homogeneous ISM environment, this is given by the relation: 
\begin{align}
\chi_{\text{ism}} & = 
266\,\Psi_{\text{ism}}\,E^{1/8}_{\gamma,\text{iso},53}\,(1+z)^{3/8} \, t_{\rm d}^{-3/8}\\
& = \left(\frac{p-0.67}{0.66\,(p+0.14)}\right)^{1/2} \left( \frac{\epsilon_\gamma}{1 - \epsilon_\gamma} \right)^{3/8} \notag \\ & \indent \cdot n_0^{-1/8}\, \gamma_{\rm m}\,\xi^{-1/2}_{\rm e},
\end{align}
where $t_{\rm d}$ is the time post-burst in days. 

As discussed in Subsection~\ref{ssec:unifiedpicture}, we inferred GRB~230815A to be exploding in a homogeneous ISM environment. 
However, for completeness, we also evaluate these parameters for the stellar wind environment. The relation for $\Psi_{\text{wind}}$ is given by: 
\begin{align}
\Psi_{\text{wind}} & = 
\left(
\frac{249.4(1+z) \nu_{\rm p} t_{\rm p}}{10^{15} d_{28}^{2} S_{\nu_{\rm p}}} \right)^{1/2}    \\
& = \frac{p-2}{0.277(p-1)} \left( \frac{p-0.69}{p+0.12} \right)^{1/2}
A_{*}^{-1/2} \epsilon_{\rm e}\,\xi^{-3/2}_{\rm e},
\end{align}
where $A_{*}$ is a parameter related to the normalisation of the density profile in a wind environment. 
In the stellar wind environment, the density is expected to fall as $n(r) = Ar^{-2}$, where $A$ is related to the mass-loss rate $\dot{M}_w$ and the progenitor's stellar wind velocity $v_w$ by $A = \dot{M}_w/4\pi v_w = 5 \times 10^{11} A_*\,\text{g}\,\text{cm}^{-1}$. 
We set $A_*$ to return the normalisation of $A$, such that $A_* = 1$ represents a typical Wolf-Rayet star with $\dot{M}_w = 10^{-5}\,M_\odot\,\text{yr}^{-1}$ and $v_w = 1\,000\,\text{km}\,\text{s}^{-1}$. 
Similarly, the relation for $\chi_{\text{wind}}$ is given by:
\begin{align}
\chi_{\text{wind}} & = 
475\,\Psi_{\text{wind}}\,E^{1/4}_{\gamma,\text{iso},53}\,(1+z)^{1/4} \, t_{\rm d}^{-1/4}\\
& = \left(\frac{p-0.69}{0.65\,(p+0.12)}\right)^{1/2} \left( \frac{\epsilon_\gamma}{1 - \epsilon_\gamma} \right)^{1/4} \notag \\ & \indent \cdot A_{*}^{-1/4}\, \gamma_{\rm m}\,\xi^{-1/2}_{\rm e}.
\label{eq:chi_wind}
\end{align}

\begin{figure*}
    \centering
    \includegraphics[width=\textwidth]{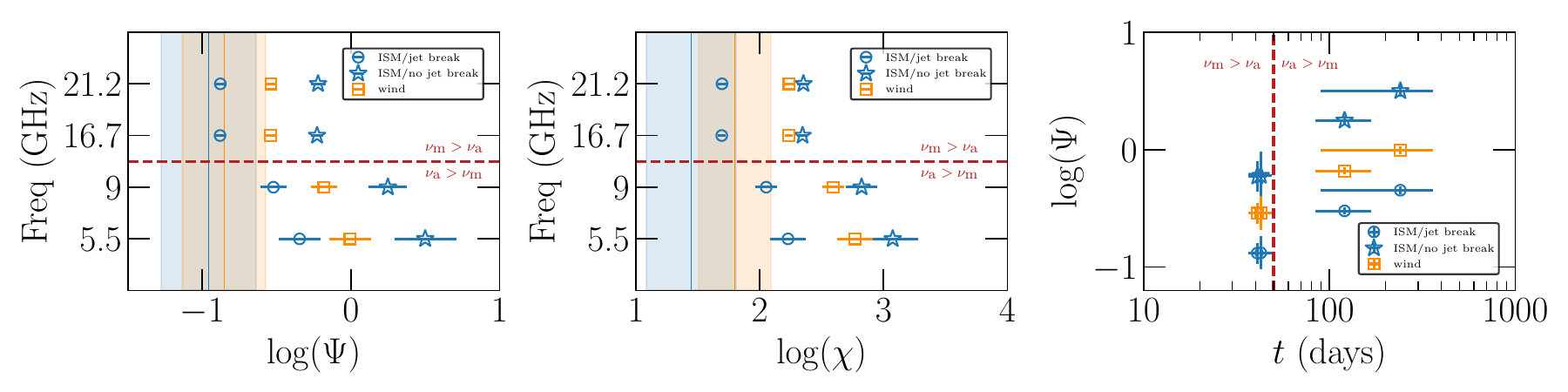}
    \caption{\textbf{\textit{Left}:} values for $\log(\Psi)$ for GRB~230815A as determined from the light-curve turnovers at 5.5, 9, 16.7, and 21.2\,GHz. 
    \textbf{\textit{Middle}:}
    values for $\log(\chi)$ for GRB~230815A as determined from the light-curve turnovers at 5.5, 9, 16.7, and 21.2\,GHz. 
    \textbf{\textit{Right}:} 
    values for $\log(\Psi)$ for GRB~230815A as a function of time in days post-burst. 
    For each panel, blue circular markers are used for measurements that assume an ISM environment and a jet break at $0.1$ days post-burst (corresponding to the X-ray jet break); blue star markers are used for measurements that assume an ISM environment and no jet break yet at $400$ days post-burst (corresponding to the observed radio behaviour); and the orange square markers are used for measurements that assume a stellar wind environment. 
    For the \textit{left} and \textit{middle} panels, the shaded regions correspond to the $1\sigma$ confidence intervals centred on the weighted averages (solid vertical lines) of the $\Psi$ and $\chi$ values, respectively, for the \cite{Duncan2023} sample in the ISM (blue) and stellar wind (orange) environments. 
    The red dashed line in each panel represents where $\nu_\textrm{m}$ passes $\nu_\textrm{a}$ (see Subsection~\ref{ssec:radiosed_new}); this is likely responsible for the apparent evolution of the parameters. 
    }
    \label{fig:psichi}
\end{figure*}

The evaluated values of $\Psi$ and $\chi$ are given in Table~\ref{tab:psichi} and also shown in Figure~\ref{fig:psichi}. 
We note here that $\chi$ is evaluated at $t_{\rm d} = 1$ day to compare with the distributions found in \citet{Duncan2023}. 
The \textit{left} and \textit{middle} panels of the figure show measurements of each parameter with their uncertainties at each frequency in both the homogeneous ISM (both assuming and not assuming the presence of a jet break at $0.1$ days post-burst) and stellar wind environments. 
These measurements are compared against the GRB population distribution determined from the \citet{Duncan2023} sample, with the $1\sigma$ confidence intervals on these parameters shaded in blue (for the ISM environment) and orange (for the stellar wind environment). 
The \textit{right} panel of the figure then shows the evolution of $\Psi$ as a function of time in days post-burst for GRB~230815A. 

\begin{table*}
\fontsize{10pt}{16pt}\selectfont 
\centering
\caption{
The $\Psi$ and $\chi$ values calculated for GRB~230815A at the observing frequencies of 5.5, 9, 16.7, and 21.2\,GHz using the turnovers in their light curves (given in parentheses), assuming they are due to the passage of $\nu_{\rm m}$.  
In the table are three sets of $\Psi$ and $\chi$ values: from left to right are the values assuming an ISM environment with the jet break occurring at $0.1$ days post-burst (corresponding to the X-ray jet break); an ISM environment with the jet break at $1\,000$ days (corresponding to the lack of jet break observed for the radio light curves assuming a two-component model); and a stellar wind environment. 
}
\label{tab:psichi}
\begin{tabular}{l|cccccc}
\toprule
 & $\Psi_{\rm ism, jet\, break}$ & $\chi_{\rm ism, jet\, break}$ & $\Psi_{\rm ism, no\, jet\, break}$ & $\chi_{\rm ism, no\, jet\, break}$ & $\Psi_{\rm wind}$ & $\chi_{\rm wind}$ \\
\hline
5.5 GHz ($241\pm151$\,days) & $0.45 \pm 0.15$ & $170 \pm 60$ & $3.2 \pm 1.5$ & $1200 \pm 600$ & $0.99 \pm 0.32$ & $\phantom{0}590 \pm 190$ \\
9.0 GHz ($121\pm47$\,days) & $0.30 \pm 0.06$ & $113 \pm 23$ & $1.8 \pm 0.5$ & $\phantom{0}660 \pm 200$ & $0.66 \pm 0.13$ & $390 \pm 80$ \\
16.7 GHz  ($41\pm5$\,days) & $0.13 \pm 0.01$ & $49 \pm 3$ & $0.59 \pm 0.06$ & $220 \pm 20$ & $0.29 \pm 0.02$ & $170 \pm 10$ \\
21.2 GHz ($41\pm6$\,days) & $0.13 \pm 0.01$ & $50 \pm 4$ & $0.60 \pm 0.07$ & $230 \pm 30$ & $0.29 \pm 0.02$ & $170 \pm 10$ \\
\hline
\end{tabular}
\end{table*}

We find that both the $\Psi$ and $\chi$ values are most consistent with the distributions found in the  \citet{Duncan2023} sample if we assume an ISM environment and a jet break at $0.1$ days corresponding to the X-ray jet break (see Figure~\ref{fig:psichi} \textit{left} and \textit{middle} panels). 
However, we stress that this analysis alone does not rule out an ISM environment with no jet break nor a wind environment. 
This is because if we consider $16.7$ and $21.2$\,GHz only, the parameter values in all three scenarios lie within the known distributions in the  \citet{Duncan2023} sample (even if not within the $1\sigma$ spread around the mean). 
As discussed in \citet{Duncan2023}, the variations in these distributions can be almost completely explained by the typical variations in the CBM density, supporting the idea of a narrow distribution (or possibly universal values) for the microphysical parameters related to the electron acceleration process, i.e., $\epsilon_{\rm e}$ and $\gamma_{\rm m}$. 

In Figure~\ref{fig:psichi} (\textit{right} panel), it is clear that there appears to be an evolution of the parameter $\Psi$ (and consequently also $\chi$) over time. 
However, in our evolving synchrotron model for our radio SEDs reported in Subsection~\ref{ssec:radiosed_new}, it is clear $\nu_\mathrm{m}$ crosses $\nu_\mathrm{a}$ at ${\sim} 50$\,days post-burst (see Figure~\ref{fig:num_nua}).  
In this scenario (when $\nu_\mathrm{a} > \nu_\mathrm{m}$ and the blast wave starting to become sub-relativistic, \textit{not} non-relativistic), the assumptions in Equations~\ref{eq:psi_ism}--\ref{eq:chi_wind} break down, and therefore we cannot interpret the increasing values of $\Psi$ with time as evidence for evolution of the microphysical parameters with time.
For the same reason, we only considered $16.7$ and $21.2$\,GHz above in the interpretation of the $\Psi$ and $\chi$ parameters in the context of their known distributions. 

Future tests of the universality or constraints on the evolution of electron acceleration processes would benefit significantly from comprehensive observations (high-cadence and long follow-up campaigns) over more frequencies (ideally above the typical self-absorption frequency of the afterglow). 
Deviations from the known distributions may point to extreme microphysical properties and provide observational evidence supporting microphysical parameters being time-variable. 
These insights are also independent of any broadband modelling, making them free from biases and variations from different modelling methodologies, while requiring significantly less multi-wavelength data and computational resources. 

\subsection{The need for a more complete and unbiased survey of radio afterglows}
\label{ssec:need4panradio}

GRB~230815A would likely not have been followed at radio frequencies without the PanRadio GRB program, and it would be even more unlikely for it to have been followed for more than a year post-burst. 
This is because radio follow-up has often been prioritised and biased toward GRBs where comprehensive multi-wavelength coverage is available \citep{Chandra2012}. 
Given the high line-of-sight extinction for this event, optical coverage was sparse and a redshift determination was not available (so it would likely not have been prioritised for radio follow-up). 

Still, even without comprehensive multi-wavelength coverage, it is clear that the multi-frequency radio light curves provided valuable insights into the system. 
As discussed in Subsection~\ref{ssec:inferences}, they were critical for providing insights into the density profile of the CBM environment, the nature of the jet structure and geometry, and constraints on dynamical and microphysical processes (in this case, the transition into the non-relativistic regime and universality of the microphysical parameters describing electron acceleration processes). 
Future events would provide even more insight if: (i) they were also triggered under the rapid-response mode (on source in $<24$\,hr); (ii) there was an even longer temporal baseline, which would be possible with more sensitive next-generation radio telescopes, to probe the non-relativistic transition and perform calorimetric measurements; and (iii) there was a redshift, which would enable a more rigorous modelling of the system and inferences of the physical properties with smaller uncertainties. 

Clearly, a complete understanding of the diversity in the GRB population and its properties would require an unbiased and comprehensive survey of radio afterglows. 
The PanRadio GRB program will extend this work to a large sample of GRBs, providing a survey of all southern \textit{Swift}-detected GRBs, following these events regardless of a detected multi-wavelength counterpart, line-of-sight extinction, or redshift measurement, etc. 
This will then provide a more unbiased view of GRB property distributions, including phenomenological properties, such as the radio brightness distribution, and physical properties, such as the jet structure. 
The larger and more unbiased sample of GRBs will also provide a better chance to follow more GRBs up to very late times. 
We should therefore be able to constrain many more GRBs through their transition to the non-relativistic regime, where calorimetric measurements can be made with minimal degeneracy with other jet parameters, such as the jet geometry. 
Additionally, we will be able to see how common a delayed jet break is at radio frequencies, whether they can be attributed to the radio-emitting electrons being farther away from the jet edge, and gain a better understanding of what determines the wind-termination-shock radius. 
Currently the primary hindrance in our understanding of these is low-sample statistics; the PanRadio GRB program aims to improve this.
A full description of the survey, the observing strategy, and initial sample analyses from bursts analysed in the first 2 years of the program will be provided in our follow-up paper (Anderson et al., \textit{in preparation}).

\section{Conclusions} \label{sec:conclusions}

GRB~230815A was the first radio afterglow to be followed from early through to very late times under the PanRadio GRB program. 
Our observing campaign for GRB~230815A lasted more than 400 days, including dedicated follow-up from ATCA and VLT/HAWK-I. 
Further optical and spectroscopic follow-up to attain a redshift was limited, however, due to the high line-of-sight extinction with $A_V=2.3$. 
We found that the early X-ray jet break at ${\sim}0.1$ days post-burst was inconsistent with the evolution of the multi-frequency radio light curves, which were approximately evolving (with some minor deviations, discussed in Subsection~\ref{ssec:unifiedpicture}) in accordance with the standard shock dynamics expected from the \citet{Blandford1976} self-similar expansion into a homogeneous ISM environment prior to a jet break. 
One option we propose to reconcile these features is to invoke a two-component jet: the early X-ray break originates from a very narrow component with half-opening angle ${\sim}2.1\degr$, while the observed evolution of the radio light curves stems from a wider component with half-opening angle $\gtrapprox 35\degr$. 

Throughout the 400-day campaign, we did not find any evidence suggesting a change in the CBM density profile or a transition to the non-relativistic regime. 
The lack of transition into the non-relativistic regime at $400$ days post-burst constrains the ratio between the blast-wave kinetic energy and the CBM density to $\log[(E_{\rm kin,iso}/n_0)/{\rm erg\ cm}^3] > 52.9$ (assuming a GRB redshift of $z=1.5$ like before), which disfavours particularly low blast-wave kinetic energies and high CBM densities. 
By tracing the evolution of the afterglow in multiple frequencies, we were also able to put some constraints on the evolution of the microphysical shock parameters describing the electron acceleration processes; in particular, we do not find any evidence supporting the evolution of these parameters with time nor any significant deviation of these parameters from their distributions derived from a wider sample of GRBs. 

Many of the insights revealed about GRB~230815A were only possible due to a multi-frequency, high-cadence campaign up to very late times about 400 days post-burst. 
The PanRadio GRB program -- a systematic, multi-year, radio survey of all southern \textit{Swift} GRB events, tracing the multi-frequency evolution of their afterglows from within an hour to years post-burst, conducted using ATCA --  will extend these insights from single events to a comprehensive sample level. 
Through the sample analysis, we will then be able to answer many of the remaining questions about long GRBs and the diversity of events within this subpopulation, from their jet structures/geometries, environments, and energetics to their dynamical and microphysical evolutions. 

\section*{Acknowledgments}
We thank the anonymous referee for the suggestions that improved the clarity of the manuscript. 
JKL acknowledges support from the University of Toronto and Hebrew University of Jerusalem through the University of Toronto--Hebrew University of Jerusalem Research and Training Alliance program. 
The Dunlap Institute is funded through an endowment established by the David Dunlap family and the University of Toronto. 
BS acknowledges the support of the French Agence Nationale de la Recherche (ANR), under grant ANR-23-CE31-0011 (project PEGaSUS). 
FDC acknowledges support from the UNAM-PAPIIT grant IN113424. 
AJG is grateful for support from the Forrest Research Foundation. 
BPG acknowledges support from STFC grant No. ST/Y002253/1 and the Leverhulme Trust grant No. RPG-2024-117. 
FS acknowledges the support of the French Agence Nationale de la Recherche (ANR), under grant ANR-22-CE31-0012 (project MOTS). 
RLCS acknowledges support from The Leverhulme Trust grant RPG-2023-240. 
NRT acknowledges support from STFC grant ST/W000857/1.
Part of this research was supported by the Australian Research Council Centre of Excellence for Gravitational Wave Discovery (OzGrav), project No. CE230100016.
The Australia Telescope Compact Array is part of the Australia Telescope National Facility (\url{https://ror.org/05qajvd42}) which is funded by the Australian Government for operation as a National Facility managed by CSIRO.
We acknowledge the Gomeroi people as the Traditional Owners of the Observatory site.
Based on observations collected at the European Southern Observatory under ESO programmes 110.24CF.014 and 110.24CF.019.
This work made use of data supplied by the UK Swift Science Data Centre at the University of Leicester. 
This research made use of {\sc Photoutils}, an {\sc Astropy} package for the detection and photometry of astronomical sources \citep{Bradley2024}.

\vspace{5mm}
\facilities{
ATCA,
\textit{Swift},
SOAR,
VLT:Yepun
}

\software{
{\sc Astropy} \citep{Astropy2013, Astropy2018}, 
{\sc Bilby} \citep{Ashton2019},
{\sc Dynesty} \citep{Speagle2020},
{\sc matplotlib} \citep{Hunter2007}, 
{\sc NumPy} \citep{Harris2020}, 
{\sc pandas} \citep{McKinney2010}, 
{\sc Photoutils} \citep{Bradley2024},
{\sc SciPy} \citep{SciPy2020}, 
and {\sc Miriad} \citep{Sault1995}.
}

\newpage

\bibliography{references}{}
\bibliographystyle{aasjournal}

\appendix
\section{Appendix: Posterior Parameter Distributions}
\label{appendix:corner}

\renewcommand\thefigure{A\arabic{figure}}    
\setcounter{figure}{0}

We provide Figures~\ref{fig:corner_xrt_sbpl}--\ref{fig:corner_semi_emp} in the appendix showing the posterior distributions of the parameters, i.e., the corner plots, for all the models we have fit in this work using our nested sampling procedures.

\begin{figure*}[h]
    \centering
    \includegraphics[width=0.95\linewidth,clip]{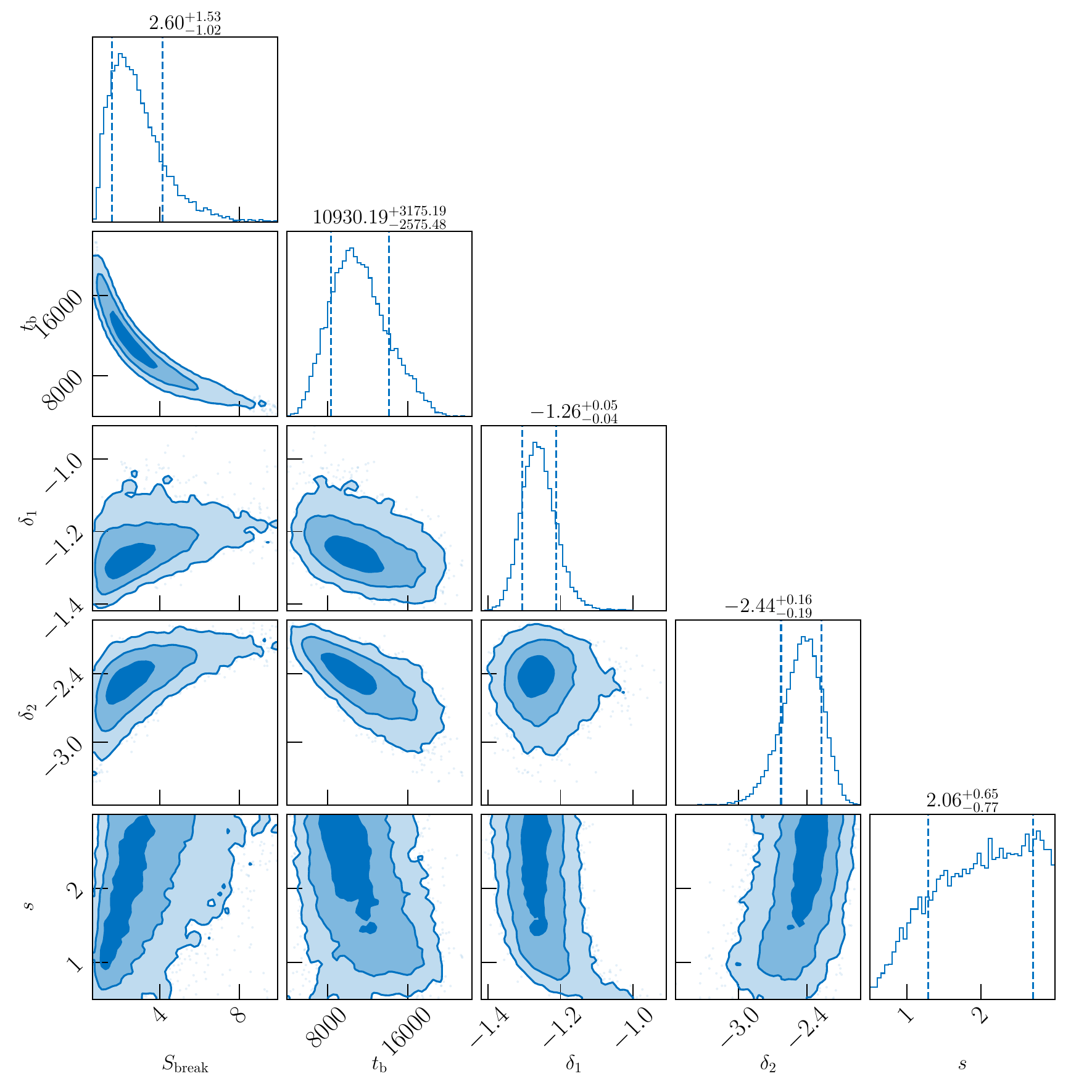}
    \caption{Posterior distributions of the parameters in the smoothly broken power-law model fit to the \textit{Swift}/XRT light curve (see Table~\ref{tab:xrtlc_fits}). The dashed vertical lines on the histograms represent the 68\% credible intervals of the corresponding marginalised posterior distributions.
    }
    \label{fig:corner_xrt_sbpl}
\end{figure*}

\begin{figure*}
    \centering
    \includegraphics[width=0.6\linewidth,clip]{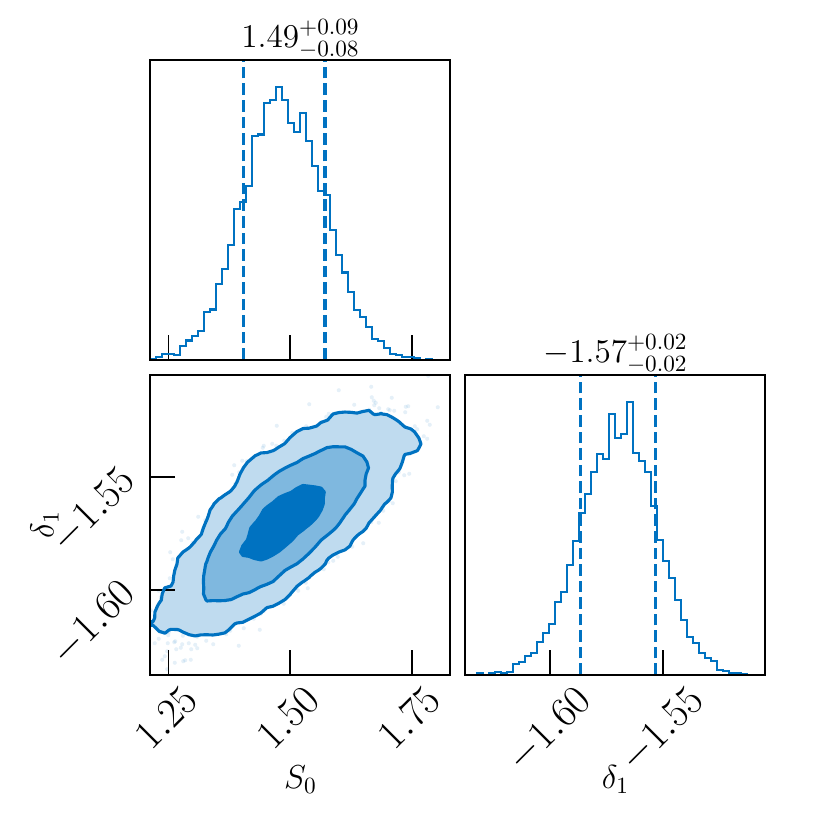}
    \caption{Similar to Figure~\ref{fig:corner_xrt_sbpl} but for the  power-law model fit of the \textit{Swift}/XRT light curve (see Table~\ref{tab:xrtlc_fits}).}
    \label{fig:corner_xrt_pl}
\end{figure*}

\begin{figure*}
    \centering
    \includegraphics[width=\linewidth,clip]{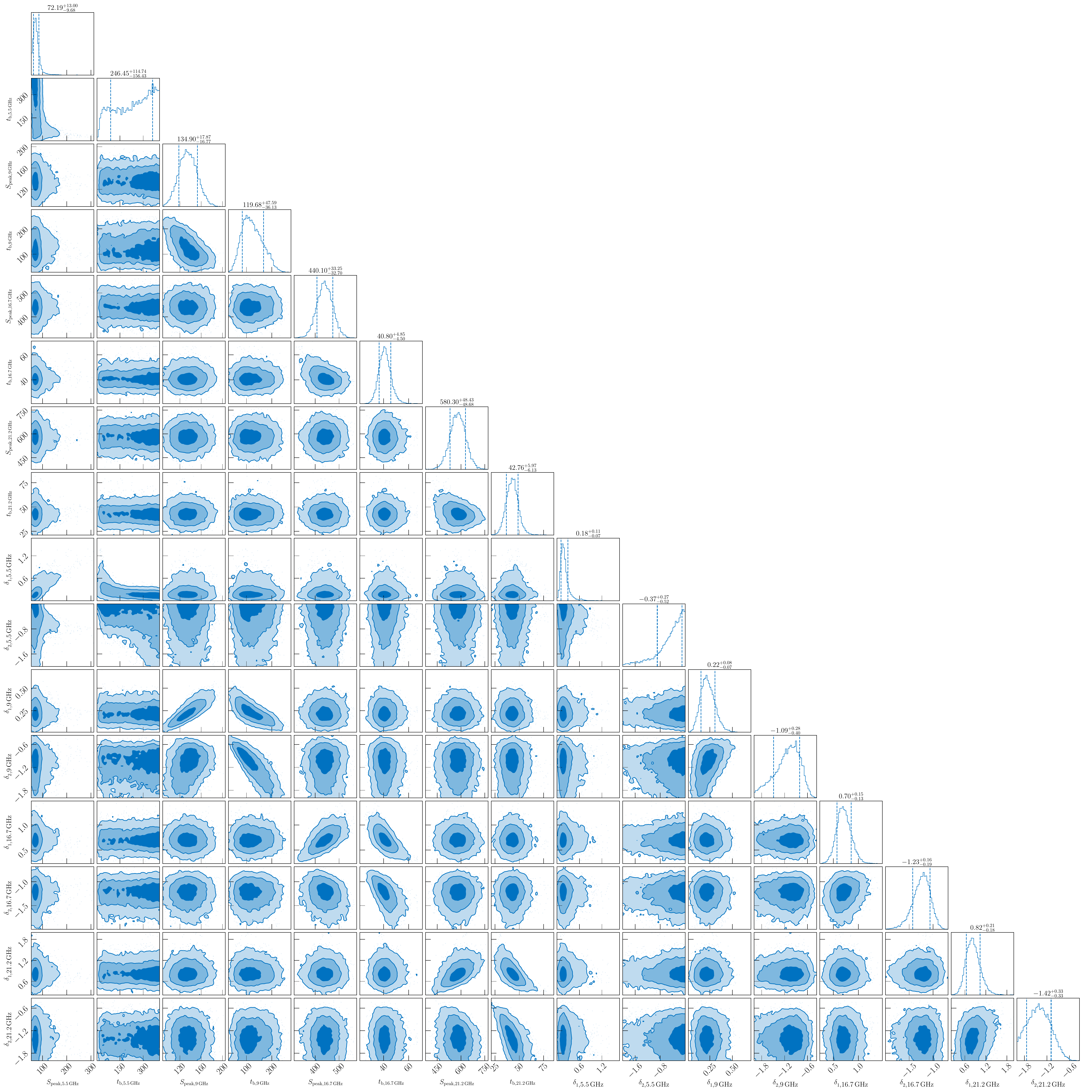}
    \caption{Posterior distributions of the parameters in the smoothly broken power-law model fit to the multi-frequency radio light curves, assuming the rise and decay slopes ($\delta_1$ and $\delta_2$) are independent for each frequency (see Table~\ref{tab:radiolc_fits}). The dashed vertical lines on the histograms represent the 68\% credible intervals of the corresponding marginalised posterior distributions. 
    To help with readability, the parameters from left to right are: $S_{\rm peak, 5.5\,GHz}$, $t_{\rm b, 5.5\,GHz}$, $S_{\rm peak, 9\,GHz}$, $t_{\rm b, 9\,GHz}$, $S_{\rm peak, 16.7\,GHz}$, $t_{\rm b, 16.7\,GHz}$, $S_{\rm peak, 21.2\,GHz}$, $t_{\rm b, 21.2\,GHz}$, $\delta_{1\textrm{,5.5\,GHz}}$, $\delta_{2\textrm{,5.5\,GHz}}$, $\delta_{1\textrm{,9\,GHz}}$, $\delta_{2\textrm{,9\,GHz}}$, $\delta_{1\textrm{,16.7\,GHz}}$, $\delta_{2\textrm{,16.7\,GHz}}$, $\delta_{1\textrm{,21.2\,GHz}}$, and $\delta_{2\textrm{,21.2\,GHz}}$.    
    }
    \label{fig:corner_radiolc_sbpl_joint}
\end{figure*}

\begin{figure*}
    \centering
    \includegraphics[width=\linewidth,clip]{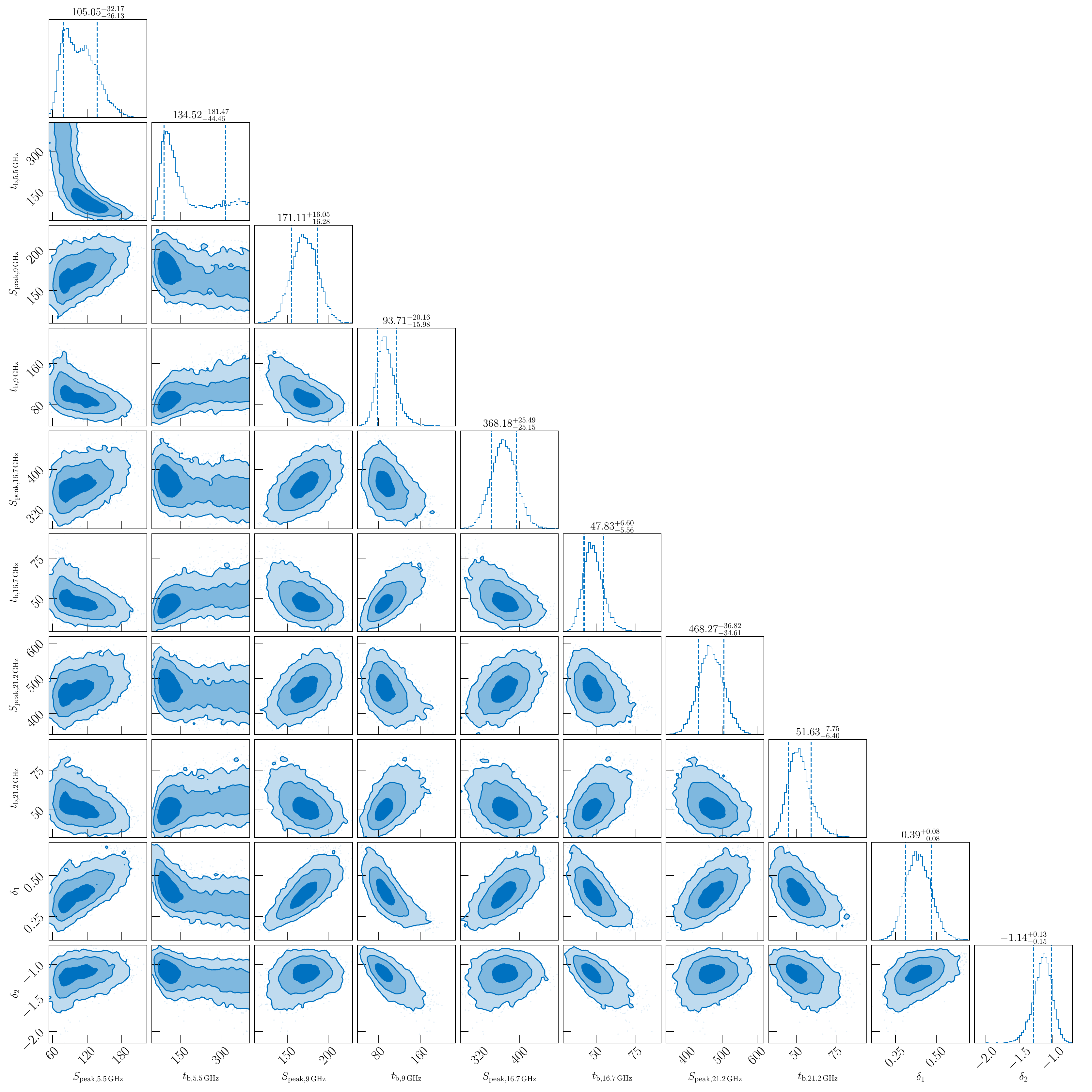}
    \caption{Similar to Figure~\ref{fig:corner_radiolc_sbpl_joint} but for the smoothly broken power-law model fit to the multi-frequency radio light curves, assuming the same rise and decay slopes ($\delta_1$ and $\delta_2$) at all frequencies (see Table~\ref{tab:radiolc_fits}).
    }
    \label{fig:corner_radiolc_sbpl_fixd1d2}
\end{figure*}

\begin{figure*}
    \centering
    \includegraphics[width=\linewidth,clip]{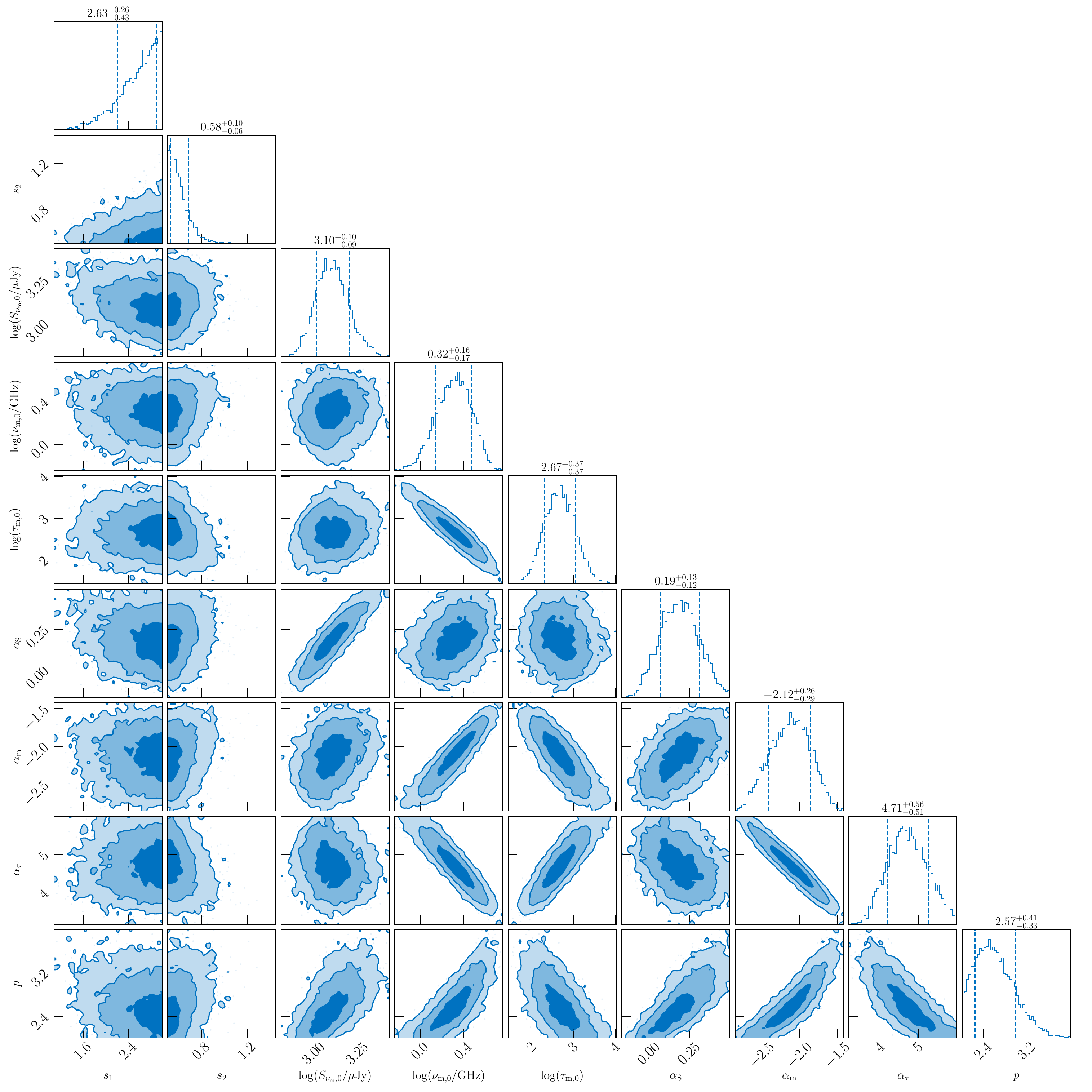}
    \caption{Posterior distributions of the parameters of the evolving synchrotron spectrum model of Subsection \ref{ssec:radiosed_new} (see Table~\ref{tab:semi_emp_fit}). The dashed vertical lines on the histograms represent the 68\% credible intervals of the corresponding marginalised posterior distributions.}
    \label{fig:corner_semi_emp}
\end{figure*}

\end{document}